\documentclass[
 preprint,
 superscriptaddress,
 preprintnumbers,
 nofootinbib,
 amsmath,amssymb,
 aps, 
 prc,
 showkeys,
 floatfix,
]{revtex4-2}

\usepackage{graphicx,
dcolumn,
bm,
xcolor,
booktabs,
mathrsfs,
dsfont,
microtype,
placeins}
\usepackage[T1]{fontenc}
\usepackage[normalem]{ulem}
\usepackage[colorlinks=true,allcolors=blue]{hyperref}

\begin{document}


\title{ShARK: A Stochastic Transport Framework for the Relativistic Relaxation Time Approximation Boltzmann Equation}

\author{Tiago Nunes da Silva}
\email{Contact author: t.j.nunes@ufsc.br}
\author{Jadna L. Barauna}
\affiliation{Departamento de Física, Universidade Federal de Santa Catarina, Florianópolis, Brazil}

\author{Giorgio Torrieri}
\affiliation{Universidade Estadual de Campinas (Unicamp), R. S\'ergio Buarque de Holanda, 777, Campinas, Brazil, 13083-859}

\date{\today}%

\begin{abstract}
We show that the Anderson-Witting relaxation-time approximation (relativistic Bhatnagar–Gross–Krook (BGK) equation) emerges as the $N\rightarrow \infty$ limit of a stochastic $N$-body particle gas, in which collisions perform a full microcanonical momentum redraw that enforces local energy-momentum conservation at every event. Using the RAMBO algorithm to sample the Lorentz-invariant phase space, we derive the finite-$N$ single-particle momentum spectrum and show that it converges to the Jüttner-Boltzmann equilibrium distribution as the local particle number grows. We couple this relaxation kernel to an advection-relaxation splitting scheme to construct ShARK (Stochastic Advection Relaxation Kinetics), a 3D Monte Carlo relativistic solver for the relaxation time approximation Boltzmann equation. The framework is conceptually related to lattice Boltzmann methods, but formulated in continuous momentum space to avoid the accuracy loss caused by finite-order momentum discretizations far from local equilibrium. We validate the numerical results against analytical solutions for two highly symmetric conformal expansions, the Bjorken and Gubser flows. The framework provides a first-principles route to far-from-equilibrium relativistic transport, with applications ranging from heavy-ion collisions to expanding astrophysical plasmas.

\end{abstract}

\maketitle
\newpage


\section{\label{sec:intro}Introduction}

Relativistic fluid dynamics provides the framework for describing bulk evolution in systems ranging from cosmological expansion and neutron star mergers to the quark-gluon plasma (QGP) created in ultrarelativistic heavy-ion collisions~\cite{deGroot:1980dk, cercignani:02relativistic, Heinz:2013th, Romatschke:2017ejr, Alford:2017rxf, Most:2021zvc}. Across these diverse disciplines, a central theoretical challenge is accurately modeling how highly anisotropic, far-from-equilibrium initial states dynamically relax into the near-equilibrium fluid regime.

Current approaches typically model this bulk evolution using second-order relativistic viscous hydrodynamics, such as the Israel-Stewart or DNMR formulations~\cite{Israel:1976tn, Denicol:2012cn}, which rely on truncating the hydrodynamic gradient expansion~\cite{Baier:2007ix, denicol2022microscopic}. While this truncation is well justified near local thermal equilibrium, the gradient series is asymptotically divergent, strictly limiting its domain of validity~\cite{Heller:2015dha}. In the presence of large spatial and temporal gradients---such as those characterizing early pre-equilibrium phases~\cite{Berges:2020fwq, Kurkela:2018vqr}, expanding dilute coronas~\cite{Werner:2013tya, Kanakubo:2021qcw}, or astrophysical shockwaves~\cite{Bouras:2009nn}---truncated fluid dynamics breaks down. Analogous truncation artifacts plague broader transport applications: state-of-the-art simulations of neutrino transport in neutron star mergers rely on moment-closure (M1-type) schemes~\cite{Foucart:2020qjb}, and standard treatments of cosmological neutrino decoupling truncate the phase-space multipole hierarchy~\cite{Froustey:2020mcq}. In kinetic theory frameworks for the pre-equilibrium dynamics of nuclear collisions, calculating response functions requires truncating an infinite hierarchy of momentum moments at a large finite order to achieve numerical convergence~\cite{Kamata:2020mka}. In these far-from-equilibrium regimes, the neglected higher-order gradients and exponentially decaying non-hydrodynamic modes become dominant, leading to unphysical artifacts, potential causality violations, and the ultimate failure of the macroscopic approximation~\cite{Florkowski:2013lya, Plumberg:2021bme}.

Relativistic kinetic theory provides a continuous description valid across arbitrary gradient scales. By tracking the microscopic one-particle phase-space distribution $f(x,p)$, the Boltzmann equation inherently resums the infinite series of macroscopic gradient corrections and captures the non-hydrodynamic modes. In particular, the relativistic Boltzmann equation in the relaxation time approximation (RTA), formulated by Anderson and Witting~\cite{Anderson:1974nyl}, provides a theoretically controlled framework:
\begin{equation}
  p^\mu \partial_\mu f = -\frac{p \cdot u}{\tau_R}(f - f_\text{eq}),
  \label{eq:RTA}
\end{equation}
where $u^\mu$ is the local fluid four-velocity, $\tau_R$ is the relaxation time controlling the transport coefficients, and $f_\text{eq}$ is the local J\"{u}ttner equilibrium distribution. Equation~\eqref{eq:RTA} admits analytical solutions~\cite{Bjorken:1982qr,Gubser:2010ze,Denicol:2014xca} that demonstrate the existence of a universal non-equilibrium kinetic attractor---a boundary-condition-independent manifold that interpolates between the free-streaming and Navier-Stokes limits, capturing the all-order dynamics where truncated hydrodynamics fails.

Solving Eq.~\eqref{eq:RTA} numerically in 3+1D without artificial truncations or numerical artifacts remains a long-standing challenge. Eulerian finite-difference and finite-volume discretizations of the spatial advection operator introduce numerical dissipation errors---a well-known limitation in relativistic fluid solvers~\cite{Rischke:1995ir}---that can artificially modify the kinetic relaxation dynamics.

The relativistic Lattice Boltzmann method (RLBM)~\cite{Mendoza:2010as, Romatschke:2011qp, Romatschke:2011hm, BAZZANINI2021101320, Gabbana:2019ydb} replaces the continuous momentum dependence with a finite set of discrete velocities and implements relaxation toward a discretized equilibrium distribution. While this provides an efficient kinetic description, its accuracy depends sensitively on the representation of the relativistic equilibrium in the discrete momentum space. Early RLBM implementations were found to depart from full kinetic calculations at high temperatures and flow velocities, a limitation attributed to the low-order approximation of the Maxwell--Jüttner distribution~\cite{Hupp:2011tz}. Subsequent formulations addressed this issue through higher-order expansions and dedicated momentum-space quadratures~\cite{Gabbana:2019ydb,Ambrus:2016fki}. Nevertheless, convergence studies of ultrarelativistic lattice Boltzmann models show that the number of discrete velocities must be substantially increased as the system approaches the ballistic regime~\cite{Ambrus:2016fki}, illustrating the computational cost associated with resolving nonequilibrium momentum distributions strongly using a fixed discrete-velocity representation.

Monte Carlo cascade algorithms~\cite{Zhang:1997ej, Zhang:1999bd,Xu:2004mz} provide a bottom-up realization of kinetic theory by representing the distribution with discrete test particles whose momenta remain continuous and whose collisions are sampled stochastically. In such approaches, macroscopic properties like collective flow and transport coefficients emerge from the underlying microscopic collision dynamics. BAMPS~\cite{Xu:2004mz,Uphoff:2014cba}, for example, has provided a fully 3+1D implementation of partonic transport and has been successfully applied to heavy-ion collisions, including studies of rapid momentum isotropization~\cite{Xu:2007aa} and the emergence of elliptic flow from microscopic interactions~\cite{Xu:2008av}. The difficulty with using a microscopic description as a general framework for heavy-ion phenomenology is not simply its computational cost, but the fact that the macroscopic properties of the medium are indirect consequences of the collision kernel. 

The deeper question underlying these approaches is whether, and under what conditions, a given numerical algorithm converges to the Anderson--Witting equation. For RLBM, the connection to relativistic kinetic theory has been established at the hydrodynamic level through Chapman--Enskog analysis~\cite{Gabbana:2019ydb}. A different question arises for particle-based algorithms: can the macroscopic collision term be derived directly from the underlying stochastic dynamics of a finite-$N$ particle system, rather than imposed phenomenologically at the level of the one-particle distribution? Establishing such a connection requires controlling the combined limits of particle number, spatial resolution, and time step. More generally, the derivation of kinetic equations from stochastic $N$-body systems remains an active mathematical problem~\cite{Butt__2023,butta2023particlesystemsbgkequation,Pulvirenti:2014}, with substantially fewer results available for relativistic equations. 

In this paper, we propose an intermediate approach that provides direct control over the transport coefficients (like hydrodynamics) while evolving in continuous momentum space (like a cascade) to avoid gradient truncations. Specifically, we formulate the collision algorithm as a stochastic $N$-body master equation. We show that an advection-relaxation operator splitting, in which particles free-stream between cells and then undergo a stochastic full redraw of momenta from the microcanonical Lorentz-Invariant Phase Space (LIPS) via the RAMBO algorithm~\cite{Kleiss:1985gy}, generates a Markov process that, in the thermodynamic large-$N$ limit, reduces exactly to the one-particle Anderson--Witting Boltzmann equation~\eqref{eq:RTA}. Because the algorithm samples continuous momentum space and imposes collisions stochastically based on the local relaxation time $\tau_R$, it intrinsically resums all hydrodynamic gradients without truncation.

We implement this framework as ShARK (Stochastic Advection Relaxation Kinetics), a general-purpose 3D Monte Carlo Particle-in-Cell (PIC)~\cite{Harlow1961tech, Harlow1964book, Hockney1981} framework. To validate the solver and demonstrate its capability to bypass truncations, we benchmark ShARK against analytical solutions of the Anderson--Witting equation. We first demonstrate reproduction of the 1D Bjorken expansion~\cite{Bjorken:1982qr}, confirming the correct emergence of viscous heating and hydrodynamization. We then perform a 3D benchmark against the Gubser flow~\cite{Gubser:2010ze, Denicol:2014xca}, utilizing a conformal mapping to a static de Sitter geometry $dS_3 \otimes \mathbb{R}$. In both geometries, we show that ShARK closely tracks the all-order non-equilibrium kinetic attractors, succeeding where second-order truncations (such as Israel-Stewart and DNMR) fail. 

This paper is organized as follows. In Sec.~\ref{sec:kinetic_theory}, we review the Anderson--Witting equation and its moments. In Sec.~\ref{sec:derivation}, we present the first-principles derivation connecting the RAMBO full-redraw algorithm to the Anderson--Witting equation through the $N$-body Master equation, and derive the finite-$N$ microcanonical single-particle spectrum. Sec.~\ref{sec:shark_framework} details the numerical implementation of ShARK. In Sec.~\ref{sec:results} we present validation results for the Bjorken and Gubser flows, along with 3D Glauber spatial convergence tests. We conclude in Sec.~\ref{sec:conclusions}. Natural units $\hbar=c=k_B=1$ are used throughout, with a mostly-plus metric signature.

\section{Relativistic Kinetic Theory in the Relaxation Time Approximation}
\label{sec:kinetic_theory}

\subsection{The one-particle distribution function}
\label{subsec:distribution}

The state of a relativistic gas is encoded in the one-particle phase-space distribution function $f(x^\mu, p^\mu)$, defined such that $f(x,p)\,d^3x\,d^3p/(2\pi)^3$ gives the mean number of particles in the phase-space volume element $d^3x\,d^3p$ around $(x^\mu, p^\mu)$ at spacetime point $x^\mu$. For massless particles, the on-shell condition $p^\mu p_\mu = 0$ constrains all momenta to the forward light cone, $p^0 = |\vec{p}|$, and the Lorentz-invariant momentum-space measure is $d^3p/p^0$.

The macroscopic fluid state is specified by the local energy-momentum tensor $T^{\mu\nu}$ and the particle four-current $N^\mu$, defined as momentum moments of $f$:
\begin{align}
    T^{\mu\nu}(x) &= \int \frac{d^3p}{(2\pi)^3 p^0}\,
        p^\mu p^\nu\, f(x,p),
    \label{eq:Tmunu} \\
    N^{\mu}(x) &= \int \frac{d^3p}{(2\pi)^3 p^0}\,
        p^\mu\, f(x,p).
    \label{eq:Nmu}
\end{align}
The local rest frame (LRF) is defined by the timelike eigenvector $u^\mu$ of $T^{\mu\nu}$, normalized as $u^\mu u_\mu = -1$,  through the Landau--Lifshitz matching condition~\cite{Landau:1987Fluid}:
\begin{equation}
    T^{\mu\nu} u_\nu = -\epsilon\, u^\mu,
    \label{eq:Landau}
\end{equation}
which defines both the fluid four-velocity $u^\mu$ and the local energy density $\epsilon$. In the conformal massless gas considered here, the equation of state is $P = \epsilon/3$. The macroscopic  energy density is related to the local temperature $T$ by the Stefan-Boltzmann law,
\begin{equation}
    \epsilon = 3P = g_{\rm eff} \frac{\pi^2}{30} T^4,
    \label{eq:stefan_boltzmann}
\end{equation}
where $g_{\rm eff}$ is the effective degeneracy factor for the relevant partonic degrees of freedom. In the numerical 
implementation and theoretical benchmarks presented in this work, we adopt normalized thermal units such that $g_{\rm eff} \pi^2 / 30 = 1$, allowing the local temperature to be extracted directly as $T = \epsilon^{1/4}$.

\subsection{The Anderson--Witting equation}
\label{subsec:AW}

The evolution of $f$ is governed by the relativistic Boltzmann equation,
\begin{equation}
    p^\mu \partial_\mu f(x,p) = \mathcal{C}[f],
    \label{eq:Boltzmann}
\end{equation}
where $\mathcal{C}[f]$ is the collision functional. The left-hand side describes free streaming: in the absence of collisions,
particle worldlines are straight ($p^\mu = \text{const}$) and $f$ is constant along them.

We adopt the relaxation time approximation (RTA) introduced by Anderson and Witting~\cite{Anderson:1974nyl}:
\begin{equation}
    \mathcal{C}[f] = \frac{p^\mu u_\mu}{\tau_R}\,
        \bigl(f - f_{\rm eq}\bigr),
    \label{eq:AW}
\end{equation}
where $\tau_R$ is the relaxation time and $f_{\rm eq}$ is the local J\"{u}ttner--Boltzmann equilibrium distribution,
\begin{equation}
    f_{\rm eq}(x,p) = \frac{g_{\rm eff}}{(2\pi)^3}\,
        \exp\!\left(-\frac{p^\mu u_\mu}{T}\right),
    \label{eq:Juttner}
\end{equation}
with $T$ the local temperature. The prefactor $p^\mu u_\mu$ in Eq.~\eqref{eq:AW} equals the negative particle energy in the LRF, $p^\mu u_\mu = -E_{\rm LRF}$, so the collision rate is proportional to the particle energy in the rest frame. This is the defining feature of the Anderson--Witting prescription, distinguishing it from the alternative Marle form~\cite{Marle:1969}, in which the relaxation rate is proportional to the particle rest mass (and therefore vanishes for massless particles). For a massless gas, the Anderson--Witting equation is the natural covariant generalization that recovers the non-relativistic Bhatnagar–Gross–Krook (BGK) equation~\cite{Bhatnagar:1954zz} in the appropriate limit~\cite{Anderson:1974nyl}.

Note that Eq.~\eqref{eq:AW} conserves energy and momentum: the collision term~\eqref{eq:AW} satisfies
\begin{equation}
    \int \frac{d^3p}{(2\pi)^3 p^0}\, p^\nu\,
        \mathcal{C}[f] = 0.
    \label{eq:collision_conservation}
\end{equation}
This follows directly from the Landau matching condition for the energy-momentum tensor together with the transversality of the shear stress. Because the Anderson--Witting kernel carries an explicit factor of $u \cdot p$, its momentum moment evaluates to the difference between the full and equilibrium energy-momentum tensors contracted with the fluid velocity:
\begin{equation}
    \int \frac{d^3p}{(2\pi)^3 p^0}\, p^\nu\, \mathcal{C}[f] = \frac{u_\mu}{\tau_R} \left( T^{\mu\nu} - T_{\rm eq}^{\mu\nu} \right) = \frac{u_\mu \pi^{\mu\nu}}{\tau_R}.
\end{equation}
By construction, the non-equilibrium dissipative fluxes are transverse to the fluid velocity ($u_\mu \pi^{\mu\nu} = 0$). Taking the $\nu$-th moment of Eq.~\eqref{eq:Boltzmann} and applying this transversality condition then yields the conservation law:
\begin{equation}
    \partial_\mu T^{\mu\nu} = 0.
    \label{eq:conservation}
\end{equation}

A central feature of the framework developed in this work is that Eq.~\eqref{eq:conservation} is satisfied at the microscopic level. When the relaxation kernel is stochastically triggered (Section~\ref{sec:shark_framework}), the framework enforces four-momentum conservation within the fluid cell. Consequently, Eq.~\eqref{eq:collision_conservation} holds for every individual relaxation realization rather than only in expectation. This microscopic enforcement circumvents the grid truncation errors that typically compromise conservation laws in Eulerian finite-difference solvers~\cite{Rischke:1995ir, Schenke:2010rr, Okamoto:2016pbc}.

It is useful to decompose $T^{\mu\nu}$ into its equilibrium and non-equilibrium parts. Using the projector $\Delta^{\mu\nu} \equiv g^{\mu\nu} + u^\mu u^\nu$ onto the space orthogonal to $u^\mu$~\cite{Israel:1976tn,deGroot:1980dk},
\begin{equation}
    T^{\mu\nu} = \epsilon\, u^\mu u^\nu
        + P\, \Delta^{\mu\nu}
        + \pi^{\mu\nu},
    \label{eq:Tmunu_decomp}
\end{equation}
where $\pi^{\mu\nu}$ is the shear stress tensor satisfying $\pi^{\mu\nu} u_\nu = 0$ and $\pi^\mu{}_\mu = 0$. In the conformal
massless gas, bulk viscosity vanishes identically, and $\pi^{\mu\nu}$ is the sole dissipative contribution to $T^{\mu\nu}$.

In ShARK, $T^{\mu\nu}$ is reconstructed cell by cell from the particle momentum distribution as
\begin{equation}
    T^{\mu\nu}\big|_{\rm cell} = \frac{1}{V_{\rm cell}}
        \sum_{i \in {\rm cell}} \frac{p_i^\mu p_i^\nu}{p_i^0},
    \label{eq:Tmunu_PIC}
\end{equation}
where the sum runs over all particles in the cell and $V_{\rm cell}$ is the cell volume. The fluid velocity $u^\mu$ and energy density $\epsilon$ are extracted from Eq.~\eqref{eq:Tmunu_PIC} via the Landau condition~\eqref{eq:Landau}, and $\pi^{\mu\nu}$ is obtained as the residual from Eq.~\eqref{eq:Tmunu_decomp}. In practice, we extract the diagonal components $T^{xx}$, $T^{yy}$, $T^{zz}$ directly from the particle sums and compute $\pi^{\mu\nu}$ in the LRF, where $u^\mu = (1,0,0,0)$ simplifies the
decomposition~\cite{Rezzolla:2013dea}.

In practice, the macroscopic quantities $u^\mu$, $\epsilon$, and $\pi^{\mu\nu}$ are dynamically reconstructed from the discrete momentum distribution of the simulated particle ensemble, as detailed in Section~\ref{subsec:reconstruction}.

\subsection{Transport coefficients from Chapman--Enskog expansion}
\label{subsec:transport}

To connect the relaxation time $\tau_R$ to the physical shear viscosity $\eta$, we use the first-order Chapman--Enskog expansion of the Anderson--Witting equation. Writing $f = f_{\rm eq} + \delta f$ with $\delta f$ first order in gradients and inserting into Eq.~\eqref{eq:AW}, one finds~\cite{Anderson:1974nyl,
deGroot:1980dk}:
\begin{equation}
    \delta f = \frac{\tau_R}{p^\mu u_\mu}\,
        p^\mu \partial_\mu f_{\rm eq} + \mathcal{O}(\partial^2).
    \label{eq:CE_deltaf}
\end{equation}
Substituting Eq.~\eqref{eq:CE_deltaf} into the definition of $\pi^{\mu\nu}$ and evaluating the momentum integrals for a
conformal massless gas gives~\cite{Anderson:1974nyl, ANDERSON1974489, Florkowski:2013lya}:
\begin{equation}
    \pi^{\mu\nu} = -2\eta\,\sigma^{\mu\nu},
    \qquad
    \eta = \frac{4}{5}\,\tau_R P,
    \label{eq:eta}
\end{equation}
where $\sigma^{\mu\nu} \equiv \tfrac{1}{2}(\nabla^\mu u^\nu + \nabla^\nu u^\mu) - \tfrac{1}{3}\Delta^{\mu\nu}(\nabla \cdot u)$ is the shear rate tensor and $\nabla^\mu \equiv \Delta^{\mu\nu} \partial_\nu$ is the spatial gradient operator in the LRF. Equivalently, using $P = \epsilon/3$, so that $\epsilon + P = (4/3)\epsilon = 4P$,
\begin{equation}
    \eta = \frac{1}{5}\,\tau_R\,(\epsilon + P).
    \label{eq:eta2}
\end{equation}
Equations~\eqref{eq:eta}--\eqref{eq:eta2} are the constitutive relations that translate a target value of $\eta/s$ into a relaxation time $\tau_R$, and thereby into the collision probability $P_{\rm trig}$ per time step used in the ShARK relaxation kernel (Section~\ref{sec:shark_framework}). For a conformal massless gas, the entropy density is $s = (\epsilon + P)/T$, so
\begin{equation}
    \frac{\eta}{s} = \frac{1}{5}\,\tau_R\,T.
    \label{eq:etas}
\end{equation}
Given a target $\eta/s$ and a local temperature $T$ extracted
from the cell, Eq.~\eqref{eq:etas} determines $\tau_R$, which in
turn sets $P_{\rm trig} \sim \Delta t / \gamma \tau_R$ at each time step, where $\gamma$ is the Lorentz factor of the local fluid rest frame relative to the stationary computational grid (laboratory frame).

Two remarks are in order. First, the Chapman--Enskog result is a first-order gradient expansion valid when $\tau_R \partial_\mu \ll 1$, i.e., in the near-equilibrium regime. The ShARK simulation does not impose this condition: cells with large gradients and small particle occupancy are accurately described by the full stochastic kinetic evolution, not by Eq.~\eqref{eq:eta}. The Chapman--Enskog relation is used only to set the target $\tau_R$ from physical input; whether the cell is actually near equilibrium is determined by the evolution and measured via $Re^{-1}$.

The relation $\tau_R = 5\eta / (\epsilon + P)$ holds for a conformal gas of massless particles. In future extensions of the ShARK framework incorporating massive degrees of freedom and realistic lattice QCD equations of state, conformal symmetry is broken. Consequently, the Chapman--Enskog matching must be extended to account for finite bulk viscosity $\zeta$, and the shear relaxation time will acquire mass-dependent thermal corrections governed by the massive J\"{u}ttner--Boltzmann distribution. For the present work, which benchmarks the algorithmic fidelity of the stochastic momentum redraws against conformal solutions, the massless relation remains sufficient.

Second, instead of relying on empirical proportionality constants or discrete linear approximations, ShARK leverages the Poissonian nature of the Anderson--Witting collision operator. The probability that a cell undergoes a stochastic momentum redraw in a given time step $\Delta t$ is evaluated as $P_{\rm trig} = 1 - \exp(-\Delta t / \gamma \tau_R)$. This ensures that the continuum relaxation rate is recovered regardless of the chosen discrete time step, provided $\Delta t \ll \tau_R$ is maintained to resolve the hydrodynamic gradients accurately. The Bjorken and Gubser benchmarks in Section~\ref{sec:results} serve to validate that this stochastic sampling yields the correct viscous transport.

\subsection{The non-equilibrium attractor and hydrodynamization}
\label{subsec:attractor}

Before describing the numerical implementation, it is useful to recall the concept of the hydrodynamic attractor, which provides the theoretical framework for interpreting our results. For systems undergoing rapid expansion, it has been established that far-from-equilibrium initial conditions converge to a universal non-equilibrium solution of the kinetic equation before local equilibrium is reached~\cite{Heller:2015dha, Behtash:2017wqg, Denicol:2018pak}. This attractor solution lies between free streaming and ideal hydrodynamics and is the kinetic-theory analog of the viscous hydrodynamic regime.

Within the Anderson--Witting framework, the attractor can be characterized quantitatively by the inverse Reynolds
number~\cite{Denicol:2012cn}:
\begin{equation}
    Re^{-1} \equiv \frac{\sqrt{\pi^{\mu\nu}\pi_{\mu\nu}}}
        {\epsilon + P},
    \label{eq:Reynolds}
\end{equation}
which measures the magnitude of shear stress relative to the enthalpy. In the LRF of a longitudinally expanding system, this reduces to $Re^{-1} = |P_L - P_T|/(\epsilon + P)$, where $P_L$ and $P_T$ are the longitudinal and transverse pressures. Traditionally, the fluid description is considered valid when $Re^{-1}$ falls below a threshold where the gradient expansion is expected to converge~\cite{Denicol:2010xn,Florkowski:2013lya}. However, a defining feature of the hydrodynamic attractor is that the system dynamically collapses onto a universal trajectory even before this convergence regime is reached. 

\section{From Stochastic Sampling to the Anderson--Witting Equation}
\label{sec:derivation}

We now establish the theoretical foundation of ShARK by demonstrating that an advection-relaxation operator splitting generates a Markov process whose continuum limit recovers the Anderson--Witting equation~\eqref{eq:AW}. This equivalence holds provided two conditions are met: the particles undergo a stochastic \textit{full redraw} of momenta from the microcanonical phase space to recover the correct equilibrium state, and the local probability of this relaxation is governed by the macroscopic relaxation time $\tau_R$ to recover the correct viscous transport. 

\subsection{The RAMBO algorithm and the microcanonical ensemble}
\label{subsec:rambo}

The RAMBO (Random Momentum BOosting) algorithm~\cite{Kleiss:1985gy} is a Monte Carlo phase-space generator originally developed for high-energy physics cross-section calculations. Given a system of $N$ massless particles with total four-momentum $P^\mu = (\sqrt{s}, \vec{0})$ in the center-of-momentum (CM) frame, RAMBO generates a new set of four-momenta $\{p_i^\mu\}_{i=1}^N$ by the following procedure:
\begin{enumerate}
    \item Generate $N$ isotropic massless four-vectors $q_i^\mu$ with energies drawn from a product of exponential distributions.
    \item Boost all $q_i^\mu$ to the frame in which their total three-momentum vanishes.
    \item Rescale all energies by a common factor $\sqrt{s}/E_{\rm CM}$ so that the total energy equals $\sqrt{s}$.
\end{enumerate}
The key mathematical property of this procedure, proven in Ref.~\cite{Kleiss:1985gy}, is that it samples each $N$-body
momentum configuration $\{p_i^\mu\}$ with equal probability: the resulting distribution is the unique uniform measure on the Lorentz-Invariant Phase Space (LIPS),
\begin{equation}
    d R_N(s) = \prod_{i=1}^N \frac{d^3 p_i}{(2\pi)^3\, 2E_i}\,
        (2\pi)^4 \delta^{(4)}\!\left(P - \sum_{i=1}^N p_i\right),
    \label{eq:LIPS}
\end{equation}
normalized to the $N$-body LIPS volume $R_N(s)$.

In statistical mechanics, the uniform measure over all $N$-body configurations consistent with fixed total energy and momentum is the microcanonical ensemble. RAMBO therefore realizes microcanonical sampling of a massless $N$-particle
gas with invariant mass $\sqrt{s}$: every accessible microscopic state is visited with equal weight.

Given that RAMBO samples the uniform LIPS measure, the marginal energy distribution of any single particle is determined by integrating out the remaining $N-1$ degrees of freedom. We derive this distribution using the recursive phase-space integration technique of Byckling and Kajantie~\cite{BycklingKajantie}.

The $N$-body LIPS volume for massless particles is~\cite{BycklingKajantie}:
\begin{equation}
    R_N(s) = \frac{(\pi/2)^{N-1}}{(N-1)!\,(N-2)!}\,s^{N-2}.
    \label{eq:RN}
\end{equation}
The single-particle invariant momentum distribution follows from the Byckling--Kajantie recursion relation:
\begin{equation}
    \frac{d^3 R_N}{d^3 p_1} = \frac{1}{2E_1}\,R_{N-1}(s'),
    \label{eq:BK_recursion}
\end{equation}
where $s' = (P - p_1)^2$ is the invariant mass squared of the
remaining $N-1$ particles. In the CM frame of the cell ($\vec{P}
= 0$, $P^0 = \sqrt{s}$) with massless particles, exact
four-momentum conservation gives:
\begin{equation}
    s' = s\left(1 - \frac{2E_1}{\sqrt{s}}\right),
    \label{eq:sprime}
\end{equation}
so that $s'$ vanishes when $E_1$ reaches its kinematic maximum $\sqrt{s}/2$. Substituting Eqs.~\eqref{eq:RN} and~\eqref{eq:sprime} into Eq.~\eqref{eq:BK_recursion} and converting to the energy distribution via $d^3p = 4\pi E_1^2\,dE_1$ (massless on-shell), the factors combine to yield the microcanonical single-particle energy spectrum:
\begin{equation}
    f_{\rm micro}(E;\,N,s) \propto
        E\left(1 - \frac{2E}{\sqrt{s}}\right)^{\!N-3},
    \qquad 0 \leq E \leq \frac{\sqrt{s}}{2}.
    \label{eq:fmicro}
\end{equation}
This result does not appear in this assembled form in the existing RAMBO~\cite{Kleiss:1985gy} or phase-space
integration~\cite{BycklingKajantie} literature, because those works target cross-section calculations weighted by matrix
elements $|\mathcal{M}|^2$ rather than the uniform phase-space measure relevant to statistical mechanics.

Two properties of Eq.~\eqref{eq:fmicro} are immediate. First, the mean energy per particle is:
\begin{equation}
    \langle E \rangle_N = \int_0^{\sqrt{s}/2} E\,f_{\rm micro}(E)\,dE
    \bigg/\int_0^{\sqrt{s}/2} f_{\rm micro}(E)\,dE
    = \frac{\sqrt{s}}{N},
    \label{eq:meanE}
\end{equation}
so that $N\langle E \rangle_N = \sqrt{s}$: total energy is conserved for all $N$. Second, defining the effective
microcanonical temperature $T_{\rm micro} \equiv \sqrt{s}/(2N)$, one has $\langle E \rangle_N = 2T_{\rm micro}$, consistent with the $\langle E \rangle = 2T$ relation for a massless relativistic gas in three spatial dimensions.

\subsubsection{Large-\texorpdfstring{$N$}{N} limit:
    convergence to J\"{u}ttner--Boltzmann}
\label{subsubsec:largeN}

For fixed $E$ and $N \to \infty$ with $T_{\rm micro} = \sqrt{s}/(2N)$ held constant, Eq.~\eqref{eq:fmicro} converges to the J\"{u}ttner--Boltzmann distribution. Expanding the logarithm:
\begin{equation}
    \ln f_{\rm micro}(E;\,N,s)
    = \ln E + (N-3)\ln\!\left(1 - \frac{2E}{\sqrt{s}}\right)
    + \text{const},
\end{equation}
and using $\ln(1-x) \approx -x$ for $x = 2E/\sqrt{s} \ll 1$ in the large-$N$ limit:
\begin{equation}
    \ln f_{\rm micro}(E;\,N,s)
    \approx \ln E - (N-3)\frac{2E}{\sqrt{s}}
    \approx \ln E - \frac{E}{T_{\rm micro}},
\end{equation}
so that
\begin{equation}
    f_{\rm micro}(E;\,N,s)
    \xrightarrow{N \to \infty}
    f_{\rm eq}(E) \propto E\,e^{-E/T_{\rm micro}}
    \equiv f_{\rm J\ddot{u}ttner}(E).
    \label{eq:Juttner_limit}
\end{equation}
The J\"{u}ttner--Boltzmann distribution therefore emerges as the thermodynamic limit of the microcanonical LIPS measure: it is not imposed externally, but arises from the underlying phase-space geometry. The leading finite-$N$ correction to this limit scales as $\mathcal{O}(1/N)$, as follows from comparing the binomial suppression $(1-2E/\sqrt{s})^{N-3}$ to the exponential approximation $e^{-E/T_{\rm micro}}$.

\subsection{The \texorpdfstring{$N$}{N}-body Markov process}
\label{subsec:Nbody}

We now describe the stochastic dynamics of the collision step and derive its governing equation. Consider a computational
cell containing $N$ particles with momenta $\{\vec{p}\} \equiv (\vec{p}_1, \ldots, \vec{p}_N)$ at time $t$. When the RAMBO kernel is triggered, it discards all current momenta and draws a completely new configuration $\{\vec{p}'\}$ from the LIPS measure~\eqref{eq:LIPS} with the same total four-momentum $P^\mu = \sum_i p_i^\mu$.

Because the new configuration $\{\vec{p}'\}$ is determined entirely by the constraints $(E_{\rm tot}, \vec{P}_{\rm tot})$ and a fresh set of random numbers, it retains no memory of the individual momenta $\{\vec{p}\}$ preceding the collision. This destruction of microscopic memory is the defining property of a Markov process. The stochastic evolution of the joint $N$-particle distribution $F_N(\{\vec{p}\}, t)$ is therefore governed by the $N$-body Master Equation:
\begin{align}
    \frac{\partial F_N(\{\vec{p}\},t)}{\partial t}
    &= \int d^{3N}p'\,\bigl[
        \mathcal{W}_N(\{\vec{p}'\} \to \{\vec{p}\})\,
        F_N(\{\vec{p}'\},t)
        \nonumber \\
    &\quad
        - \mathcal{W}_N(\{\vec{p}\} \to \{\vec{p}'\})\,
        F_N(\{\vec{p}\},t)
    \bigr],
    \label{eq:master_N}
\end{align}
where $\mathcal{W}_N(\{\vec{p}\} \to \{\vec{p}'\})$ is the joint $N$-body transition rate generated by a single RAMBO call. Its form follows directly from the uniform LIPS measure:
\begin{equation}
    \mathcal{W}_N(\{\vec{p}\} \to \{\vec{p}'\}) =
    \frac{1}{\tau_R}\,\frac{dR_N(s)}
        {\displaystyle\int dR_N(s)},
    \label{eq:WN}
\end{equation}
where $\tau_R$ is the relaxation time setting the collision rate and $s = P^\mu P_\mu$ is the cell invariant mass. Equation \eqref{eq:WN} states that all outgoing configurations $\{\vec{p}'\}$ consistent with the conservation constraints are reached with equal probability, which is the microcanonical statement.

To connect the $N$-body dynamics to the one-particle kinetic theory of Sec.~\ref{sec:kinetic_theory}, we seek the evolution of the one-particle distribution
\begin{equation}
    f_1(\vec{p}_1, t) \equiv
    \int d^3p_2 \cdots d^3p_N\,
    F_N(\vec{p}_1, \ldots, \vec{p}_N,\, t),
    \label{eq:f1_def}
\end{equation}
obtained by integrating out the $N-1$ unobserved momenta. Marginalizing Eq.~\eqref{eq:master_N} over $d^3p_2\cdots d^3p_N$ introduces a coupling between $f_1$ and the two-particle distribution $F_2$, and more generally produces the relativistic BBGKY hierarchy. To close this hierarchy at the one-particle level, we invoke the molecular chaos hypothesis, that before each collision step, the momenta of the $N$ particles within a cell are statistically independent. The joint distribution therefore factorizes:
\begin{equation}
    F_N(\vec{p}_1, \ldots, \vec{p}_N, t)
    \simeq \prod_{i=1}^N f_1(\vec{p}_i, t).
    \label{eq:chaos}
\end{equation}

This assumption is the standard closure used in deriving the Boltzmann equation from BBGKY, adapted to the present discrete-time Markov setting. Its physical content is that inter-particle correlations built up during a single RAMBO
event are destroyed before the next collision step, either by the free-streaming advection phase, which transports particles from different cells to different positions, or by the subsequent redraw, which replaces all momenta simultaneously. Mathematically, this decoupling is governed by the \textit{propagation of chaos} property~\cite{McKean:1967,Sznitman:1991}. In a standard stochastic mean-field limit where $N \to \infty$ and the interaction strength scales as $1/N$, the factorization in Eq.~\eqref{eq:chaos} is dynamically preserved at leading order. In the context of the ShARK algorithm, the primary source of inter-particle correlation arises from the microcanonical momentum redraws dictated by Eq.~\eqref{eq:WN}: energy-momentum conservation requires that changing the momentum of any single particle inevitably restricts the available phase space for all other particles in the local computational cell. However, because this kinematic recoil is democratically shared among the remaining $N-1$ particles, the resulting correlations scale as $\mathcal{O}(1/N)$. Consequently, in the dense thermodynamic limit ($N \gg 1$), these finite-$N$ conservation correlations are suppressed. The individual particle trajectories factorize, the discrete microcanonical ensemble approaches the continuous canonical J\"{u}ttner--Boltzmann equilibrium state, and the system recovers the one-particle continuum kinetic limit in accordance with the molecular chaos hypothesis.

While a fully rigorous propagation-of-chaos result for the Anderson--Witting equation remains outstanding, researchers have extensively studied the non-relativistic BGK equation in this context. Notably, Butt\`a, Hauray, and Pulvirenti~\cite{Butt__2021, Butt__2023} formally proved that the discrete stochastic particle system converges to the continuum BGK solution in the large-$N$ limit using an analogous framework. Their proof establishes that the stochastic momentum redraws can be mathematically driven by continuous fields evaluated within localized spatial neighborhoods. This provides a theoretical foundation for ShARK's discrete grid-based fluid reconstruction (Phase~2) and stochastic momentum sampling (Phase~3). The derivation presented here provides the physicist's argument that the relativistic case follows the same structure; numerical validation via the Bjorken and Gubser benchmarks (Sec.~\ref{sec:results}) confirms that the resulting continuum equation matches those analytical solutions.

Under the assumption of molecular chaos, the marginalization of Eq.~\eqref{eq:master_N} closes at the one-particle level. For any given particle (dropping the subscript 1 for brevity), the $N-1$ surrounding particles act as an uncorrelated thermal bath, and the $N$-body transition rate $\mathcal{W}_N$ marginalizes to an effective one-body rate $W(\vec{p} \to \vec{p}')$. The one-particle Master Equation is:
\begin{equation}
    \frac{\partial f(\vec{p},t)}{\partial t}
    = \int d^3p'\,\bigl[
        W(\vec{p}' \to \vec{p})\,f(\vec{p}',t)
        - W(\vec{p} \to \vec{p}')\,f(\vec{p},t)
    \bigr].
    \label{eq:master_1}
\end{equation}

\subsection{Anderson--Witting emergence}
\label{subsec:fullredraw}

In a full-redraw collision, the new momentum $\vec{p}'$ is drawn from the equilibrium distribution of the available phase space, with a rate independent of the initial momentum $\vec{p}$. The transition rate therefore factorizes as:
\begin{equation}
    W(\vec{p} \to \vec{p}') = \Gamma(\vec{p}'),
    \label{eq:fullredraw}
\end{equation}
where $\Gamma(\vec{p}')$ is the rate at which the medium populates the target state $\vec{p}'$. This captures the mathematical mechanism of RAMBO: the outgoing momentum is drawn entirely from the phase-space geometry of the new
configuration, with no memory of the initial state.

Substituting Eq.~\eqref{eq:fullredraw} into Eq.~\eqref{eq:master_1}:
\begin{align}
    \frac{\partial f(\vec{p},t)}{\partial t}
    &= \Gamma(\vec{p})\int d^3p'\,f(\vec{p}',t)
     - f(\vec{p},t)\int d^3p'\,\Gamma(\vec{p}').
    \label{eq:master_factored}
\end{align}
The two remaining integrals have clear physical interpretations:
\begin{align}
    \int d^3p'\,f(\vec{p}',t) &= n(t),
    \label{eq:n_def}\\
    \int d^3p'\,\Gamma(\vec{p}') &\equiv \frac{1}{\tau_R},
    \label{eq:tauR_def}
\end{align}
where $n(t)$ is the local particle number density and $\tau_R$ is the relaxation time, defined as the inverse total scattering rate. Substituting:
\begin{equation}
    \frac{\partial f(\vec{p},t)}{\partial t}
    = n(t)\,\Gamma(\vec{p}) - \frac{f(\vec{p},t)}{\tau_R}.
    \label{eq:master_reduced}
\end{equation}
The stationary solution of Eq.~\eqref{eq:master_reduced}
satisfies $\partial_t f = 0$, giving $\Gamma(\vec{p}) \propto f_{\rm eq}(\vec{p})$. The normalization is fixed by Eq.~\eqref{eq:tauR_def}:
\begin{equation}
    \Gamma(\vec{p}) = \frac{f_{\rm eq}(\vec{p})}{n\,\tau_R}.
    \label{eq:Gamma}
\end{equation}
Because the RAMBO algorithm explicitly conserves particle number during every stochastic momentum redraw (analogous to the energy conservation detailed in Eq.~\eqref{eq:meanE}), the local fluid density $n(t)$ remains constant over the collision step. Consequently, substituting the transition rate from Eq.~\eqref{eq:Gamma} into Eq.~\eqref{eq:master_reduced} allows the density normalization factors to cancel directly, yielding the familiar relaxation-time form:
\begin{equation}
    \frac{\partial f(\vec{p},t)}{\partial t}
    = -\frac{1}{\tau_R}\bigl(f(\vec{p},t) - f_{\rm eq}(\vec{p})\bigr).
    \label{eq:BGK_rest}
\end{equation}

This is the BGK equation in the local rest frame of the fluid cell. The full-redraw hypothesis~\eqref{eq:fullredraw} is both necessary and sufficient for this result; it uniquely produces a linear relaxation toward $f_{\rm eq}$ with rate $1/\tau_R$.

To write the kinetic equation in a Lorentz-covariant form valid in arbitrary frames, we include the free-streaming term on the left-hand side and promote the time derivative to a covariant derivative along particle worldlines. The left-hand side of the Boltzmann equation, $p^\mu \partial_\mu f$, reduces to $p^0\,\partial_t f$ in the rest frame, so Eq.~\eqref{eq:BGK_rest} becomes $p^0\,\partial_t f = -(p^0/\tau_R)(f - f_{\rm eq})$.

In a frame moving with four-velocity $u^\mu$ relative to the cell, $p^0$ in the rest frame corresponds to $-p^\mu u_\mu$ (the particle energy in the local rest frame, taken positive). The unique Lorentz-covariant generalization that reduces to Eq.~\eqref{eq:BGK_rest} in the rest frame is therefore:
\begin{equation}
    p^\mu \partial_\mu f
    = \frac{p^\mu u_\mu}{\tau_R}\,(f - f_{\rm eq}),
    \label{eq:AW_derived}
\end{equation}
which is precisely the Anderson--Witting equation~\eqref{eq:AW}~\cite{Anderson:1974nyl}. As established in Section~\ref{subsec:AW}, the defining feature of this formulation is that the invariant collision term scales proportionally with the local rest frame energy ($-p^\mu u_\mu$). This ensures that when the Boltzmann equation is evaluated in the rest frame to extract the physical time evolution ($\partial_t f = \mathcal{C}[f]/p^0$), the energy factors cancel. The resulting physical relaxation rate is exactly $1/\tau_R$ for all particles, matching the uniform, energy-independent stochastic redraw executed by the RAMBO algorithm. 

In contrast, the alternative Marle form~\cite{Marle:1969} scales the invariant collision term by the particle rest mass $m$. When evaluated in the rest frame, this yields a physical relaxation rate proportional to $m/p^0$, which would unphysically suppress the thermalization of highly energetic particles and vanish for the massless conformal gas simulated here. The Anderson--Witting prescription is therefore the covariant extension that preserves the correct ultra-relativistic transport limits~\cite{Anderson:1974nyl} while remaining consistent with the microcanonical phase-space redraw.

\subsection{Three dynamical regimes and the finite-\texorpdfstring{$N$}{N}
    interpolation}
\label{subsec:regimes}

The derivation above assumed $N \gg 1$, which justified the molecular chaos hypothesis and the large-$N$ convergence $f_{\rm micro} \to f_{\rm eq}$. In a heavy-ion collision, the local particle occupancy decreases as the system expands and dilutes. The microcanonical spectrum~\eqref{eq:fmicro} reveals how the relaxation kernel responds to this dilution without any external switching criterion.

For $N \gg 1$ (specifically $N \gtrsim 30$ in practice), the microcanonical  spectrum~\eqref{eq:fmicro} is well approximated by $f_{\rm eq}$, molecular chaos holds to $\mathcal{O}(1/N)$, and the coarse-grained evolution approaches the deterministic Anderson--Witting equation~\eqref{eq:AW}. This is the \textit{hydrodynamic regime}, in which Chapman--Enskog transport coefficients and the attractor solution emerge.

For $2 \lesssim N \lesssim 30$, finite-particle fluctuations become significant. The microcanonical spectrum deviates measurably from $f_{\rm eq}$: the power-law suppression $(1-2E/\sqrt{s})^{N-3}$ imposes a kinematic cutoff at $E = \sqrt{s}/2$ that becomes increasingly prominent as $N$ decreases, and the ensemble equivalence between microcanonical and canonical descriptions breaks down at order $1/N$ corrections. The system remains in energy-momentum conserving stochastic relaxation, but can no longer be accurately described by a deterministic fluid equation. This is the \textit{finite-$N$ kinetic regime}, in which ShARK continues to evolve the system correctly through the microcanonical dynamics, without requiring an external criterion to switch from a fluid to a kinetic description. 

By naturally shifting from the mean-field continuum limit ($N \gg 1$) in the dense core to a granular, microcanonically constrained discrete cascade at the dilute edges, ShARK intrinsically mimics the physical breakdown of the fluid approximation. This transition avoids the unphysical propagation of continuous fractional densities in the vacuum limit, effectively allowing the finite-$N$ artifacts to act as a physical proxy for the discrete nature of  freeze-out.

For $N < 2$, the cell contains fewer than two particles, and the RAMBO relaxation kernel cannot be invoked (at least two particles are required for a meaningful phase-space redistribution). Collisions therefore cease, and the dynamics reduces to free streaming along straight worldlines. This directly resolves the ballistic regime where Relativistic Lattice Boltzmann Models struggle to maintain accuracy without introducing computationally expensive discrete-velocity expansions~\cite{Ambrus:2016fki}. Consequently, ShARK is capable of smoothly transitioning from hydrodynamics to kinetic freeze-out dynamically, bypassing the need for an ad-hoc particlization surface.

The three regimes are summarized in Table~\ref{tab:regimes}. The boundary between them is set dynamically by the local particle occupancy, which in ShARK is controlled by the parameter $\alpha$ (number of test particles per unit entropy density per cell volume), the cell size $\Delta x$, and the local energy density $\epsilon$. The occupancy in a given cell at time $\tau$ is:
\begin{equation}
    N_{\rm cell}(\tau) \approx
    \alpha \cdot \frac{4}{3}\,\epsilon(\tau)^{3/4}
    \cdot V_{\rm cell},
    \label{eq:N_cell}
\end{equation}
where the factor $(4/3)\epsilon^{3/4}$ is the entropy density of the conformal massless gas. As $\epsilon$ decreases with expansion, $N_{\rm cell}$ decreases continuously, and the system transitions through the three regimes without any externally imposed switching surface.

\begin{table}[t]
\centering
\caption{Three dynamical regimes emerging from the finite-$N$
microcanonical spectrum~\eqref{eq:fmicro}. The boundaries
$N \approx 30$ and $N = 2$ are estimates; the crossover is
continuous.}
\label{tab:regimes}
\begin{ruledtabular}
\begin{tabular}{lccc}
    Regime & $N$ range & Distribution & Description \\
    \hline
    Hydrodynamic   & $N \gg 1$ ($\gtrsim 30$)
                   & $f_{\rm micro} \approx f_{\rm J\ddot{u}ttner}$
                   & BGK / Chapman-Enskog \\
    Finite-$N$ kinetic & $2 \lesssim N \lesssim 30$
                   & $f_{\rm micro}$ [Eq.~\eqref{eq:fmicro}]
                   & Stochastic relaxation \\
    Ballistic      & $N < 2$
                   & ---
                   & Free streaming \\
\end{tabular}
\end{ruledtabular}
\end{table}

\section{Numerical Implementation: The ShARK Framework}
\label{sec:shark_framework}

The theoretical framework of Sec.~\ref{sec:derivation} is implemented in ShARK as a particle-in-cell (PIC) solver on a static Eulerian Cartesian grid. The time evolution follows a first-order Lie-Trotter operator splitting~\cite{Trotter:1959ytf}, which alternates between a free-streaming advection step and a localized stochastic relaxation step. While this sequential splitting introduces a systematic truncation error of $\mathcal{O}(\Delta t)$, it avoids the doubled computational cost of the stochastic relaxation kernel required by a symmetric second-order Strang splitting scheme~\cite{Strang1968}. 

Following initialization (Sec.~\ref{subsec:grid_setup}), one complete time step $\Delta t$ consists of the following operations, executed in order:
\begin{enumerate}
    \item \textbf{Advection} (Sec.~\ref{subsec:advection}): Each test particle is propagated along its free-streaming worldline for time $\Delta t$.
    \item \textbf{Macroscopic reconstruction} (Sec.~\ref{subsec:reconstruction}): Particles are assigned to grid cells; the fluid four-velocity $u^\mu$, energy density $\epsilon$, and shear stress $\pi^{\mu\nu}$ are extracted from the particle sums.
    \item \textbf{Relaxation} (Sec.~\ref{subsec:rambo_algo}): For each cell, a RAMBO full-redraw is triggered stochastically utilizing the Poisson probability $P_{\rm trig} = 1 - \exp(-\Delta t / \gamma \tau_R)$ determined by the local $\eta/s$ and $T$.
\end{enumerate}
The kinetic decoupling criterion ($N < 2$) that naturally emerges at late times is detailed in Sec.~\ref{subsec:freezeout}.

\subsection{Grid setup and test-particle initialization}
\label{subsec:grid_setup}

The spatial domain is discretized into a static Eulerian Cartesian grid with cells of volume $V_{\rm cell} = \Delta x\,\Delta y\, \Delta z$. ShARK employs the test-particle method~\cite{PhysRevC.25.1460}, representing the one-particle distribution $f(x,p)$ by a discrete ensemble of massless computational particles. Each test particle carries a four-momentum $p_i^\mu$ and a continuous spatial position $\vec{x}_i$ that is not constrained to the grid.

The number of test particles initialized in a cell is proportional to the local entropy it contains. For a conformal massless gas, the entropy density is $s = \frac{4}{3}\epsilon^{3/4}$, so the expected particle count is
\begin{equation}
    \langle N_{\rm cell} \rangle = \lambda
    = \alpha \cdot s \cdot V_{\rm cell}
    = \frac{4\alpha}{3}\,\epsilon^{3/4}\,V_{\rm cell}.
    \label{eq:lambda}
\end{equation}
The integer count is drawn from a Poisson distribution, $N_{\rm cell} \sim \mathrm{Poisson}(\lambda)$, so that the particle number fluctuates event by event around $\lambda$ while the mean is determined by the local thermodynamic state. We draw initial momenta from the RAMBO kernel (Sec.~\ref{subsec:rambo_algo}) using the cell's invariant mass and fluid velocity from the initial condition.

The parameter $\alpha$ thus sets the number of test particles per unit entropy in each cell. Sec.~\ref{subsec:freezeout} explores its physical interpretation and role in kinetic decoupling.

\subsection{Particle advection}
\label{subsec:advection}

In the advection step, each test particle moves along its free-streaming worldline:
\begin{equation}
    \vec{x}_i(t + \Delta t) = \vec{x}_i(t)
    + \frac{\vec{p}_i}{p_i^0}\,\Delta t.
    \label{eq:advection}
\end{equation}
Particles carry continuous spatial coordinates and are not constrained to discrete grid vertices, ensuring this step is evaluated to machine precision without introducing numerical diffusion. The discrete time step $\Delta t$ must be chosen small enough that particles do not cross multiple cells per step on average. For a grid with uniform spatial spacing $\Delta x$ and massless particles ($v=1$), this imposes the kinematic constraint $\Delta t \lesssim \Delta x$. This serves as the PIC analog to the Courant--Friedrichs--Lewy (CFL) condition~\cite{Courant:1967cfl}; however, unlike in finite-difference fluid solvers where it dictates mathematical stability, here it functions as a resolution constraint to ensure accurate nearest-grid-point (NGP) binning.

ShARK employs open boundary conditions. Particles that advect beyond the defined spatial grid boundary are excluded from the subsequent grid-binning routine. Consequently, these escaped particles do not contribute to the macroscopic fluid reconstruction and bypass the relaxation step entirely. Physically, this models matter escaping into the vacuum beyond the fireball edge, as these kinetically decoupled particles continue to free-stream indefinitely without further interaction.

\subsection{Macroscopic reconstruction}
\label{subsec:reconstruction}

After advection, particles are assigned to grid cells by their continuous positions (nearest-grid-point binning). To extract the macroscopic fluid state, the full $4 \times 4$ energy-momentum tensor is first reconstructed cell by cell from the discrete particle sums:
\begin{equation}
    T^{\mu\nu}_{\rm cell} = \frac{1}{V_{\rm cell}}
    \sum_{i \in \rm cell} \frac{p_i^\mu p_i^\nu}{p_i^0}.
    \label{eq:Tmunu_cell}
\end{equation}
In general 3D Cartesian simulations, finding the Landau frame is a non-linear eigenvalue problem for extracting the fluid velocity $u^\mu$ and energy density $\epsilon$ from $T^{\mu\nu}$~\cite{Rezzolla:2013dea}. To solve this for arbitrary non-equilibrium momentum fluxes, ShARK constructs $T^\mu_{\phantom{\mu}\nu}$ as a $4 \times 4$ matrix and evaluates its spectrum at each time step using a direct numerical eigensolver. The local energy density $\epsilon$ is identified as the magnitude of the unique, strictly real timelike eigenvalue, and the associated normalized eigenvector defines the fluid four-velocity $u^\mu$.

For the highly symmetric conformal expansions evaluated in the validation benchmarks of this work (Sec.~\ref{sec:results}), the geometry of the expansion analytically prescribes the fluid velocity profile. In these test cases, we use the analytical flow directly to evaluate the local rest frame kinematics, avoiding numerical diagonalization. Once the local rest frame is established, we extract the residual shear stress $\pi^{\mu\nu}$ by projecting the discrete particle sums onto the subspace orthogonal to $u^\mu$ via Eq.~\eqref{eq:Tmunu_decomp}.

\subsection{The relaxation kernel}
\label{subsec:rambo_algo}

The localized relaxation step is implemented via the RAMBO algorithm~\cite{Kleiss:1985gy}. When triggered, RAMBO performs a full redraw of the $N_{\rm cell}$ test particles within a cell, replacing all momenta with a new configuration sampled uniformly from the $N$-body LIPS with total four-momentum equal to the cell's current $P^\mu_{\rm cell}$. The pre-collision momenta are discarded entirely, satisfying the Markovian full-redraw requirement of Sec.~\ref{subsec:fullredraw}.

The cell's invariant mass entering the RAMBO step is computed from the current particle sums immediately before the trigger:
\begin{equation}
    M_{\rm cell} = \sqrt{E_{\rm cell}^2 - |\vec{P}_{\rm cell}|^2},
    \qquad
    \vec{v}_{\rm cell} = \frac{\vec{P}_{\rm cell}}{E_{\rm cell}},
    \label{eq:M_BGK}
\end{equation}
where $E_{\rm cell} = \sum_i p_i^0$ and $\vec{P}_{\rm cell} = \sum_i \vec{p}_i$ are summed over the $N_{\rm cell}$ particles currently in the cell. 

\begin{table}[t]
\caption{RAMBO full-redraw algorithm for a fluid cell with invariant mass $M_{\rm cell}$ and fluid velocity $\vec{v}_{\rm cell}$.}
\label{alg:rambo}
\begin{ruledtabular}
\begin{tabular}{cl}
Step & Operation \\
\hline
1 & For each of the $N$ particles, sample an isotropic massless \\
  & four-vector $q_i^\mu$: direction uniform on $S^2$, energy \\
  & $E_i = -\ln(\xi_1\xi_2)$ with $\xi_{1,2}\sim\mathcal{U}(0,1]$. \\[4pt]
2 & Compute $Q^\mu = \sum_i q_i^\mu$. Boost all $q_i^\mu$ by \\
  & $\vec{\beta} = -\vec{Q}/Q^0$ to the CM frame; compute \\
  & $M_Q = \sum_i |\vec{q}_i^{\,\prime}|$. \\[4pt]
3 & Rescale: $q_i^{\prime\mu} \to q_i^{\prime\mu} \times M_{\rm cell}/M_Q$. \\
  & This enforces $\sum_i E_i^{\rm new} = M_{\rm cell}$. \\[4pt]
4 & Boost all rescaled momenta by $\vec{v}_{\rm cell}$ to the \\
  & laboratory frame. Output: $\{p_i^\mu\}$ with $\sum_i p_i^\mu = P^\mu_{\rm cell}$.
\end{tabular}
\end{ruledtabular}
\end{table}

The rate at which the RAMBO kernel is invoked sets the effective shear viscosity. From the Chapman-Enskog result (Eq.~\ref{eq:etas}), the required proper relaxation time is:
\begin{equation}
    \tau_R = 5\left(\frac{\eta}{s}\right)\frac{1}{T},
    \label{eq:tauR_cell}
\end{equation}
where the local temperature $T = \epsilon^{1/4}$ is obtained from the rest-frame energy density:
\begin{equation}
    \epsilon = \frac{M_{\rm cell}}{\gamma_{\rm cell}\,V_{\rm cell}},
    \qquad
    \gamma_{\rm cell} = \frac{1}{\sqrt{1 - |\vec{v}_{\rm cell}|^2}}.
    \label{eq:epsilon_cell}
\end{equation}

Since the simulation runs on a fixed laboratory-frame time step $\Delta t$, the proper relaxation time $\tau_R$ must be time-dilated to the laboratory frame, yielding $\tau_R^{\rm lab} = \gamma_{\rm cell} \tau_R$. Thus, in a given step $\Delta t$, the continuous Anderson--Witting equation predicts a fractional relaxation of $1 - \exp(-\Delta t/\tau_R^{\rm lab})$ toward $f_{\rm eq}$. The RAMBO kernel is then triggered with the Poisson probability:
\begin{equation}
    P_{\rm trig} = 1 - \exp\!\left(
        -\frac{\Delta t}{\gamma_{\rm cell}\,\tau_R}
    \right).
    \label{eq:Ptrig}
\end{equation}
A uniform random number $\xi \in [0,1)$ is drawn; if $\xi \leq P_{\rm trig}$, all $N$ particles in the cell undergo the RAMBO redraw. Note that in the fireball interior, $\vert{}\vec{v}_{\rm cell}\vert{} \ll 1$, so $\gamma_{\rm cell} \approx 1$ and the dilation correction is negligible; in the dilute outer corona where the bulk velocity is large, the factor $\gamma_{\rm cell}$ increases, suppressing $P_{\rm trig}$ relative to the rest-frame rate.

To prevent singularities in $\gamma_{\rm cell}$ for near-vacuum cells, the quantity $1 - |\vec{v}_{\rm cell}|^2$ is bounded to a minimum of $10^{-6}$. Additionally, to prevent division by zero in the relaxation-time calculation, we enforce a temperature floor of $T \ge 10^{-9}$. Finally, cells containing fewer than two particles, or with an invariant mass below a threshold $M_{\rm cutoff}$, are excluded from the collision step. This guarantees pairwise (or higher) momentum conservation and suppresses the amplification of statistical fluctuations in near-empty spatial regions.

\subsection{The oversampling factor and kinetic decoupling}
\label{subsec:freezeout}

A natural baseline would be to associate one test particle with one physical parton. However, for a classical massless Boltzmann gas, the entropy-to-number ratio is $s/n = 4$, so the physical parton number density is $n_{\rm phys} = s/4$ (corresponding to $\alpha_{\rm phys} = 1/4$). For the cell sizes required for spatial resolution ($\Delta x \sim 0.1$~fm), even the physical QCD system contains fewer than one parton per cell at temperatures near the QCD crossover: for a 2-flavor QGP at $T_c = 155$~MeV, the parton number density is $n \approx 1.4$~fm$^{-3}$, giving $n\,V_{\rm cell} \approx 0.001$ for $\Delta x = 0.1$~fm. Therefore, anchoring $\alpha$ to the physical parton number would be both impractical and conceptually ill-defined.

We interpret $\alpha$ as an entropy-based oversampling factor, so that the physically meaningful quantity is the dimensionless product $\alpha\,V_{\rm cell}$, which controls the number of test particles per entropy quantum $s\,\Delta x^3$ in a cell. This product determines where in energy density the ShARK dynamical regimes of Sec.~\ref{subsec:regimes} occur. Specifically, the $N < 2$ ballistic transition occurs when
\begin{equation}
    \frac{4\alpha}{3}\,\epsilon_{\rm dec}^{3/4}\,V_{\rm cell} = 2,
    \label{eq:eps_dec}
\end{equation}
which defines a kinetic decoupling energy density $\epsilon_{\rm dec}$ analogous to the particlization temperature $T_{\rm part}$ in hybrid models. In such models, $T_{\rm part} \in [135, 155]$~MeV is a free parameter (extracted via Bayesian studies~\cite{Bernhard:2019bmu}) that controls the global temperature at which the Cooper-Frye conversion is applied. In ShARK, $\alpha\,V_{\rm cell}$ is the analogous free parameter that controls the local energy density at which each cell transitions to ballistic dynamics. The two parameters carry the same physical information, but $T_{\rm part}$ enforces the transition globally across a hypersurface, whereas $\alpha\,V_{\rm cell}$ enforces it locally and independently in each cell as it dilutes.

Inverting Eq.~\eqref{eq:eps_dec}, the value of $\alpha$ corresponding to a target decoupling temperature $T_{\rm dec}$ (using $\epsilon_{\rm dec} = T_{\rm dec}^4$ for the conformal gas with the normalization of Sec.~\ref{sec:kinetic_theory}) is:
\begin{equation}
    \alpha(T_{\rm dec}) =
    \frac{3}{2\,V_{\rm cell}\,T_{\rm dec}^3}.
    \label{eq:alpha_Tdec}
\end{equation}
For $\Delta x = 0.1$~fm and $T_{\rm dec} \in [135, 150]$~MeV, Eq.~\eqref{eq:alpha_Tdec} gives $\alpha \in [3417, 4682]$. 

At late times, cells in the dilute outer corona lose particles through advection faster than collisions can replenish them. When the particle count in a computational cell drops below $N = 2$, the RAMBO algorithm cannot simultaneously conserve energy and three-momentum; consequently, the collision probability is set to zero. Furthermore, cells whose total invariant mass falls below a predefined cutoff threshold ($M_{\rm cutoff}$) are similarly excluded to prevent the amplification of statistical fluctuations. These dilute cells then evolve by free streaming alone, entering the ballistic regime of Table~\ref{tab:regimes} without requiring any ad hoc external switching criteria.

This kinetic decoupling is both local and asynchronous: each cell decouples independently when its own macroscopic and occupancy conditions drop below the threshold, rather than at a global system time or on a global freeze-out hypersurface. Hot, dense cells at the core decouple later than dilute peripheral cells, and regions with steeper expansion gradients decouple earlier than static regions. This local, geometry-sensitive decoupling serves as the microscopic realization of the continuous core-corona transition.

\section{Results}
\label{sec:results}

In this section, we present the validation results for the ShARK framework. Although ShARK is engineered as a fully 3D numerical solver, validating viscous relativistic transport algorithms requires benchmarks with known analytical solutions. Because no such solutions exist for arbitrary asymmetric 3D expansions, we impose highly symmetric initial conditions that constrain the simulated dynamics to lower dimensions. This methodology allows us to quantify the numerical fidelity of the continuous advection and discrete collision algorithms against fluid limits before deploying the model in realistic 3D collision geometries.

Furthermore, simulating infinitely expanding systems like the Bjorken and Gubser flows on a finite 3D Cartesian grid inherently introduces boundary artifacts. To circumvent this limitation, we use a coordinate mapping technique---often used in parton cascade models like BAMPS~\cite{El:2007vg, El:2008yy}---that maps the expanding spacetime into a static computational box. In this approach, the volume expansion is implemented via a continuous redshift of the particle momenta during the advection phase. This preserves the symmetries of the analytical solutions without finite-size grid effects.

To establish the accuracy of our numerical implementation, we proceed in stages. First, in Sec.~\ref{subsec:micro_validation}, we isolate the RAMBO relaxation kernel to verify its reproduction of microcanonical statistical mechanics and its dependence on the particle number $N$. Second, in Sec.~\ref{subsec:bjorken}, we benchmark the model against the $0+1$D Bjorken expansion and verify the recovery of the temperature evolution in the ideal limit. Third, in Sec.~\ref{subsec:gubser_flow}, we simulate the $0$D Gubser flow in de Sitter space to test the conformal mapping of large transverse hydrodynamic gradients. Finally, in Sec.~\ref{subsec:optical_glauber} we present results with realistic 3D initial conditions and use them for a self-convergence study.

\subsection{Microcanonical Phase Space Validation}
\label{subsec:micro_validation}

To verify the exactness of the momentum-space relaxation kernel, we isolate a single fluid cell at rest with a fixed invariant mass $M = 10$~GeV and perform $10^5$ independent stochastic redraws for discrete test-particle multiplicities $N \in \{4, 10, 100\}$. The cell's invariant mass equals the center-of-mass energy of the local $N$-particle system ($M \equiv \sqrt{s}$). The microcanonical distribution previously introduced in Eq.~\eqref{eq:fmicro} can be normalized by integrating the proportional relation over the kinematically allowed energy range $0 \leq E \leq M/2$ to obtain the normalized single-particle energy probability density:
\begin{equation}
    f_{\rm micro}(E) = \frac{4(N-1)(N-2)}{M^2} \, E \left( 1 - \frac{2E}{M} \right)^{N-3}, \quad 0 \leq E \leq M/2.
\end{equation}
In the thermodynamic limit ($N \to \infty$, $M \to \infty$, with $M/N$ fixed), this bounded microcanonical spectrum converges to the continuous canonical Jüttner distribution for a massless LIPS gas:
\begin{equation}
    f_{\rm cano}(E) = \frac{1}{T^2} \, E \exp\!\left(-\frac{E}{T}\right),
\end{equation}
characterized by an effective cell temperature $T = M/(2N)$. 

\begin{figure}[t]
    \centering
    \includegraphics[width=\linewidth]{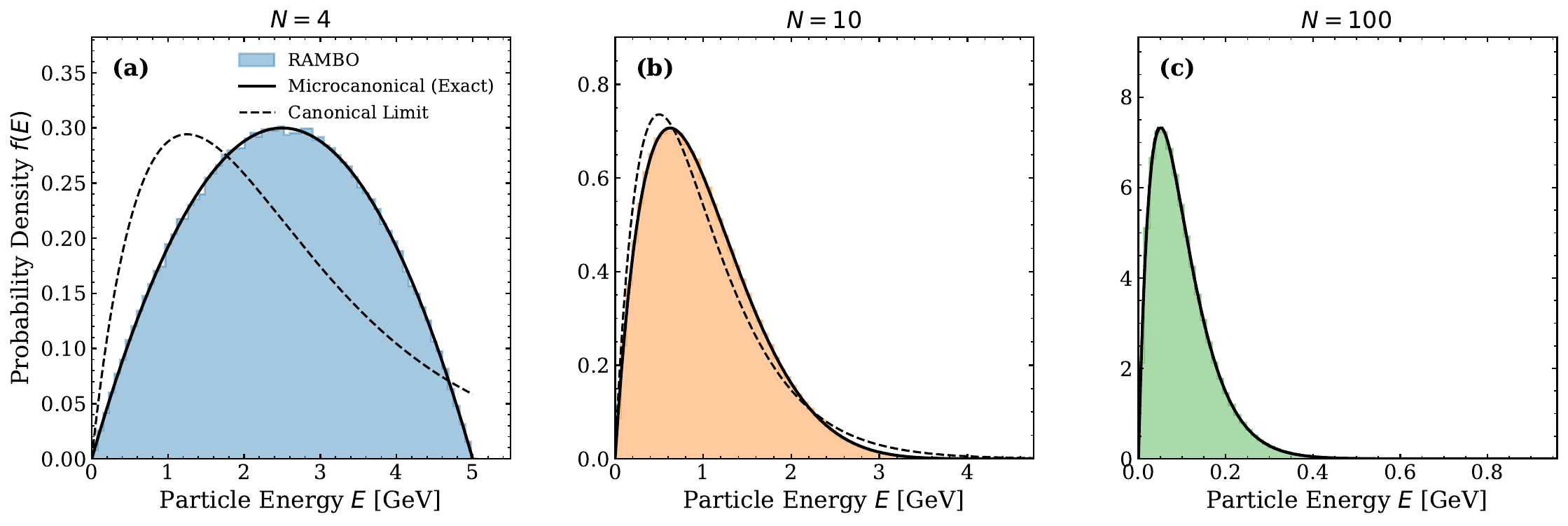}
    \caption{Single-particle energy spectra generated by the ShARK relaxation kernel for a cell with fixed invariant mass $M = 10$~GeV and test-particle multiplicities $N = 4, 10$, and $100$. The numerical histograms (blue, orange, green) are compared to the analytical microcanonical LIPS distributions (solid black lines) and the canonical Jüttner distribution (dashed black lines).}
    \label{fig:rambo_validation}
\end{figure}

In Fig.~\ref{fig:rambo_validation}, we compare the energy spectra generated by the ShARK relaxation kernel against these analytical limits. The numerical histograms agree with the microcanonical predictions across all tested particle multiplicities. For sparsely populated cells, such as $N=4$, exact energy conservation strictly limits the maximum energy of any single particle to $E = M/2$. This hard kinematic bound suppresses the high-energy tail, resulting in visible deviations from the exponential canonical limit. However, as the particle multiplicity increases, these microcanonical constraints relax, and the ensemble-averaged mean distribution rapidly converges to the continuous canonical Jüttner curve.

Although the ensemble-averaged spectrum converges quickly, it is crucial to distinguish the mean distribution from the statistical variance of a single simulated event. In a full $3+1$D transport simulation, a fluid cell undergoes only one stochastic RAMBO redraw per time step. Because relative statistical fluctuations scale as $1/\sqrt{N}$, relying on small discrete multiplicities introduces significant noise into the reconstructed stress tensor $T^{\mu\nu}$. Consequently, while the microcanonical phase-space constraints vanish for moderate $N$, the stochastic noise inherent to the Monte Carlo method requires larger particle multiplicities in practice to suppress unphysical viscous gradients and ensure smooth hydrodynamization. Sec.~\ref{subsec:optical_glauber} quantitatively explores the statistical requirements necessary to reach this continuous fluid limit.

These results confirm that the ShARK relaxation kernel captures the finite-$N$ kinetic phase-space dynamics on an event-by-event basis, while recovering the continuous thermodynamic limit required for asymptotic fluid behavior.

\subsection{Bjorken Flow}
\label{subsec:bjorken}

Following the validation of the relaxation kernel, we turn 
to the bulk fluid evolution by simulating a $0+1$D transversely 
homogeneous Bjorken expansion.

The longitudinal boost invariance of the Bjorken expansion~\cite{Bjorken:1982qr} is difficult to maintain on a finite $3$D Cartesian Eulerian grid, as particles free-streaming out of the longitudinal boundaries generate unphysical edge effects that propagate inward. To isolate the localized RAMBO relaxation kernel from spatial advection artifacts, we simulate the Bjorken expansion using the homogeneous box technique developed for the BAMPS parton cascade~\cite{El:2007vg, El:2008yy}.

In this approach, the fluid is macroscopically at rest within a static, periodic Eulerian grid ($v_x = v_y = v_z = 0$). The spatial density remains uniform and the spatial advection operator is entirely bypassed. The longitudinal expansion of the metric is treated geometrically in momentum space: during the advection step over a time interval $\Delta t$, the longitudinal momentum of every test particle is continuously redshifted according to the proper-time scale factor:
\begin{equation}
    p_z(t + \Delta t) = p_z(t) \frac{\tau}{\tau + \Delta t}.
\end{equation}
The particle energy is subsequently updated to preserve the massless 
dispersion relation, $E = \sqrt{p_x^2 + p_y^2 + p_z^2}$. This 
momentum deformation reproduces the macroscopic $\partial_\mu u^\mu = 1/\tau$ expansion rate without moving the particles spatially, guaranteeing exact boost invariance. The local energy density $\epsilon$ used to evaluate the relaxation time $\tau_R$ is scaled by the volumetric expansion factor $\tau_0/\tau$, to account for the geometric dilution of the fluid.

\begin{figure}[t]
    \centering
    \includegraphics[width=\linewidth]{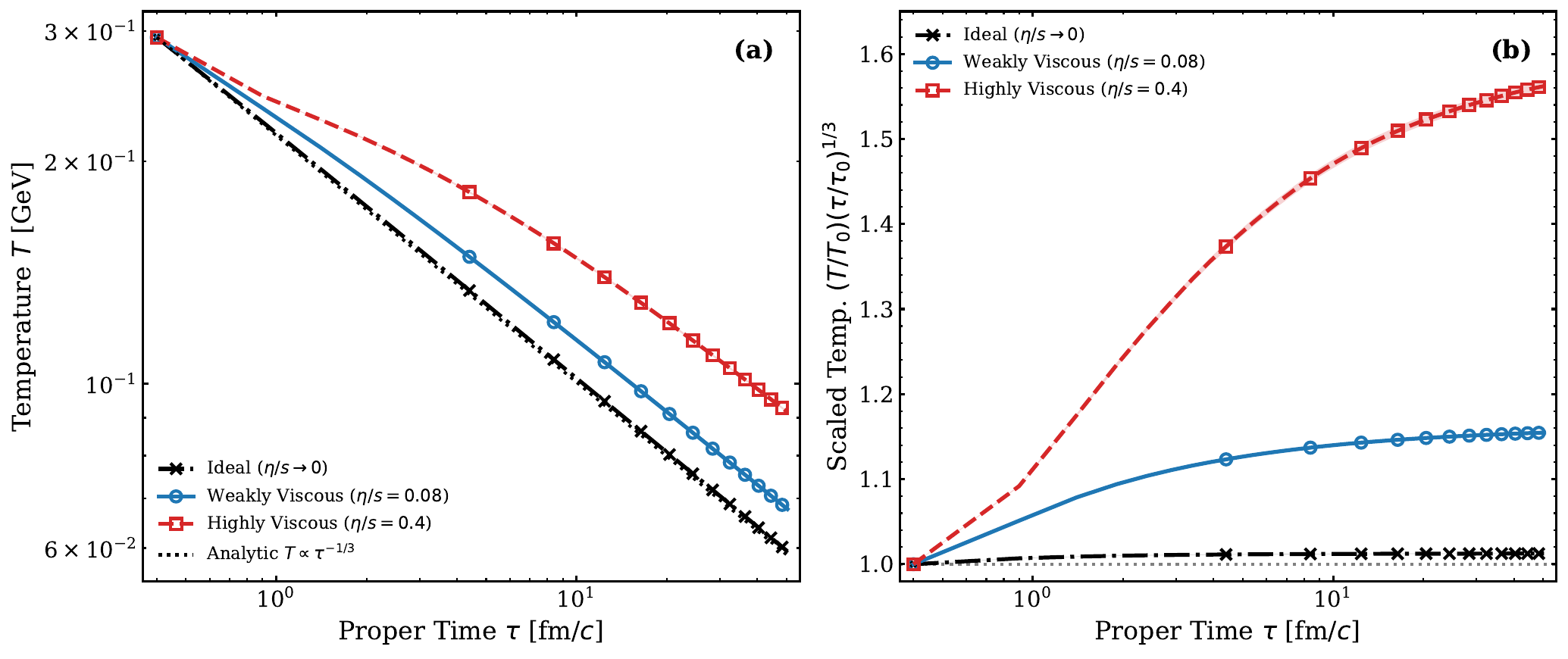}
    \caption{Temperature evolution in a $0+1$D Bjorken expansion, initialized with isotropic momenta ($P_L = P_T$). \textbf{Left:} Absolute temperature $T(\tau)$ as a function of proper time on a log-log scale. ShARK simulations are shown for near-ideal (black dashed-dotted line, $\eta/s = 10^{-4}$) and viscous runs with $\eta/s = 0.08$ (solid blue) and $0.4$ (dashed red) and compared to the analytic curve $T \propto \tau^{-1/3}$ (black dotted line). \textbf{Right:} The scaled temperature ratio $(T/T_0)(\tau/\tau_0)^{1/3}$.}
    \label{fig:bjorken_heating}
\end{figure}

We perform $0+1$D Bjorken simulations on a $10 \times 10 \times 10$ Cartesian grid with cell spacing $\Delta x = 0.2$~fm and time step $\Delta t = 0.01$~fm/$c$. The simulations are initialized at a starting proper time of $\tau_0 = 0.4$~fm/$c$. We use a uniform initial energy density of $\epsilon_0 = 100.0$~GeV/fm$^3$, setting the initial temperature $T_0$ via the conformal equation of state. While this sets a physical QGP temperature scale, reflected directly in the absolute cooling trajectories of Fig.~\ref{fig:bjorken_heating}, the conformal invariance of the massless RTA equation makes this initialization scale immaterial to the remaining non-equilibrium observables, which we evaluate via dimensionless ratios or normalize to their initial values. Since we use the homogeneous box technique, we initialize the fluid at rest ($v_x = v_y = v_z = 0$). We choose an oversampling factor of $\alpha = 30000$ to place the system into the dense thermodynamic limit ($N_{\rm cell} \gg 1$) and suppress finite-$N$ stochastic fluctuations.

We first present the system's temperature evolution against the analytical cooling rate in the ideal-fluid limit. A fundamental consequence of non-equilibrium dynamics is entropy generation. In the context of the $0+1$D Bjorken expansion, shear viscosity 
hinders the rapid longitudinal expansion of the fluid, effectively performing work against the flow and transforming collective kinetic energy into thermal energy. As a result, the fluid cools more slowly than an ideal fluid. For a conformal ideal fluid ($\eta/s \to 0$), the Bjorken cooling rate follows the power-law $T(\tau) \propto \tau^{-1/3}$. 

In Fig.~\ref{fig:bjorken_heating}, we present the temperature evolution for isotropic initial conditions ($P_L = P_T$) across three different transport regimes: the near-ideal limit ($\eta/s = 10^{-4}$), a weakly viscous fluid ($\eta/s = 0.08$), and a highly viscous fluid ($\eta/s = 0.4$). The isotropic initialization ensures that the observed heating stems entirely from the steady-state expansion rather than the transient relaxation of an initial pressure anisotropy. 

The left panel of Fig.~\ref{fig:bjorken_heating} shows the absolute temperature on a logarithmic scale. The near-ideal ShARK simulation approaches the analytic $T \propto \tau^{-1/3}$ baseline across orders of magnitude in proper time. Conversely, the runs with finite shear viscosity cool more slowly, diverging upwards from the ideal trajectory.

To quantitatively isolate this dissipative effect, the right panel of 
Fig.~\ref{fig:bjorken_heating} presents the scaled temperature ratio,  $(T/T_0)(\tau/\tau_0)^{1/3}$. The numerical results show that as $\eta/s$ increases, the system deviates monotonically upward from the ideal baseline. 

Although the $0$D simulation bypasses the spatial advection operator, the stochastic relaxation kernel remains fully active. During each time step, the geometric redshift drains $p_z$, driving the system away from equilibrium (towards $P_L \to 0$). At the same time, the stochastic trigger continuously redraws cell momenta, forcing instantaneous local isotropy ($P_L = P_T$). This dynamical competition between geometric expansion and local isotropization forces the fluid onto a universal viscous trajectory. Without the active relaxation kernel (i.e., in the free-streaming limit), the expansion redshift would simply 
drive the pressure anisotropy to $P_L = 0$ indefinitely, and the temperature would decay faster than the ideal fluid limit of $\tau^{-1/3}$.

\begin{figure}[t]
    \centering
    \includegraphics[width=.85\linewidth]{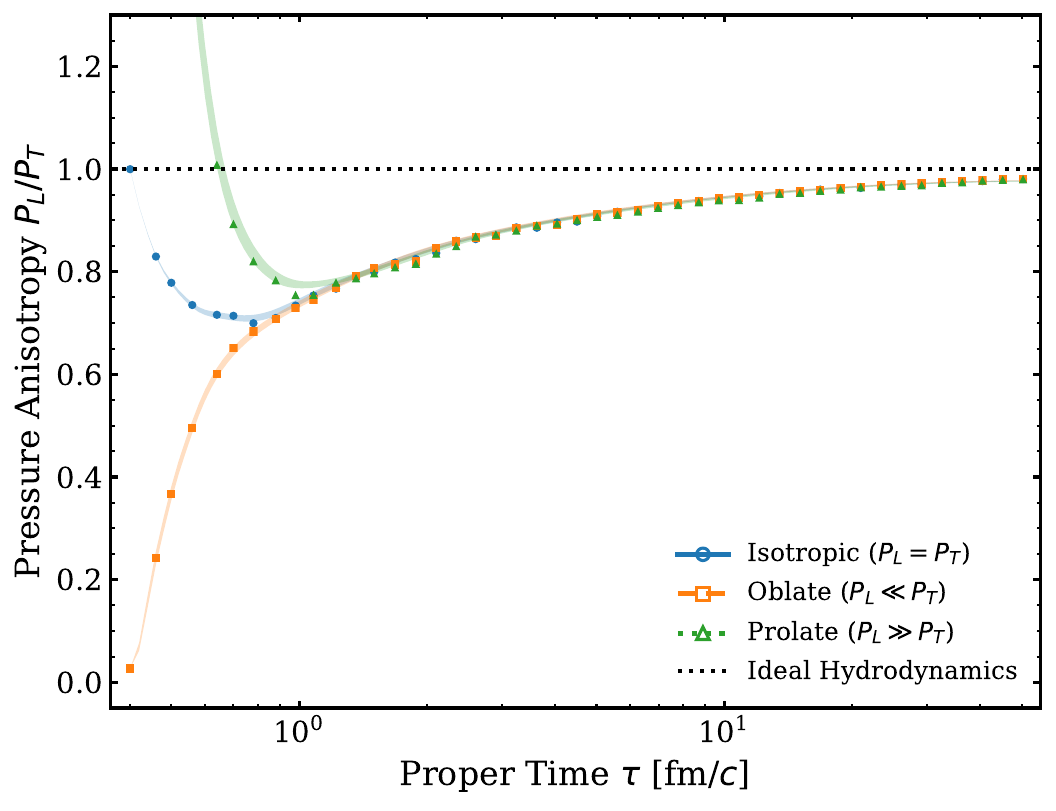}
    \caption{Proper time evolution of the pressure anisotropy $P_L / P_T$ in a $0+1$D Bjorken expanding system for a fixed shear viscosity-to-entropy density ratio of $\eta/s = 0.08$ for three distinct momentum-space initializations: an isotropic state ($P_L = P_T$, solid blue line), an oblate configuration ($P_L \ll P_T$, dashed orange line), and a prolate configuration ($P_L \gg P_T$, dotted green line). The discrete markers denote the expectation values from the Monte Carlo simulation. The shaded bands are a smoothed trajectory obtained via an adaptive log-time Gaussian kernel, with the vertical width indicating the $\pm 1\sigma$ standard error of the mean.}
    \label{fig:bjorken_anisotropy}
\end{figure}

After verifying that the relaxation kernel sources the correct viscous heating for an isotropic expansion, a natural next question is whether the fluid also loses memory of an initial pressure anisotropy, a characteristic feature of hydrodynamization. In Fig.~\ref{fig:bjorken_anisotropy}, we show the proper-time ($\tau$) evolution of the pressure anisotropy for a fluid with a fixed shear viscosity of $\eta/s = 0.08$. We initialize the simulation in three different momentum configurations: an isotropic state ($P_L = P_T$), an oblate deformation ($P_L \ll P_T$), and a prolate deformation ($P_L \gg P_T$). We generated the anisotropic initial states by deforming the isotropic particle momenta sampled by the RAMBO kernel at $\tau_0$. For the oblate case, the longitudinal momentum $p_z$ of each particle is reduced to $20\%$ of its original value. The transverse momenta $p_x$ and $p_y$ are uniformly scaled to conserve the cell's original invariant mass. In the prolate case, the transverse momenta $p_x$ and $p_y$ are reduced to $50\%$ of their original values, and $p_z$ is scaled up to conserve the local rest-frame energy.

Despite the different non-equilibrium starting conditions, the relaxation kernel drives the stress tensor toward isotropy. Within the first $10$~fm/$c$, the disparate trajectories converge onto a single viscous trajectory, reflecting the fluid's loss of memory of the initial state. At later times, the system asymptotically approaches the ideal hydrodynamic limit of $P_L / P_T = 1$, with the residual anisotropy dictated by the longitudinal expansion rate and the fixed shear viscosity.

This rapid convergence of disparate, far-from-equilibrium initial states onto a universal, initial-condition-independent trajectory, before local thermal equilibrium is reached, is the hallmark of a non-equilibrium hydrodynamic attractor~\cite{Heller:2015dha}: while the hydrodynamic gradient series is asymptotically divergent, it can be Borel-resummed to reveal a unique, non-analytic attractor solution, so that arbitrary initial data evolves towards this solution, driven by the decay of non-hydrodynamic modes.  This mathematical framework was subsequently expanded \cite{Romatschke:2017vte} by identifying analogous far-from-equilibrium attractors across diverse microscopic theories, including kinetic theory under the relaxation time approximation.  While the aforementioned convergence of the pressure anisotropy $P_L/P_T$ illustrates memory loss qualitatively, a more quantitative validation of the kinetic framework comes from mapping the system's evolution in terms of a universally scaled dimensionless variable. It is customary to evaluate the dynamics against the dimensionless conformal time $\tilde{w} \equiv \tau T / (5\eta/s)$, which measures proper time in units of the microscopic relaxation time, and captures the direct competition between the longitudinal expansion rate and the collision rate. By casting the evolution as a function of $\tilde{w}$, the disparate transient trajectories of the fluid collapse onto the universal attractor manifold \cite{Heller:2015dha}.

To verify whether the hydrodynamization process in the ShARK framework scales correctly with transport coefficients, we evaluate the scaled shear stress from the simulated energy-momentum tensor. Because the simulation executes the Bjorken expansion within a static coordinate box, we must correct the simulated energy density by a geometric volume expansion factor, $\epsilon = \epsilon_{\rm grid} (\tau_0/\tau)$. For the massless conformal fluid in our simulations, the scaled shear stress was extracted from
\begin{equation}
    \bar{\pi}(\tau) = \frac{\frac{1}{3}\epsilon - T^{zz}}{\frac{4}{3}\epsilon}.
\end{equation}

\begin{figure}[t]\centering\includegraphics[width=.85\linewidth]{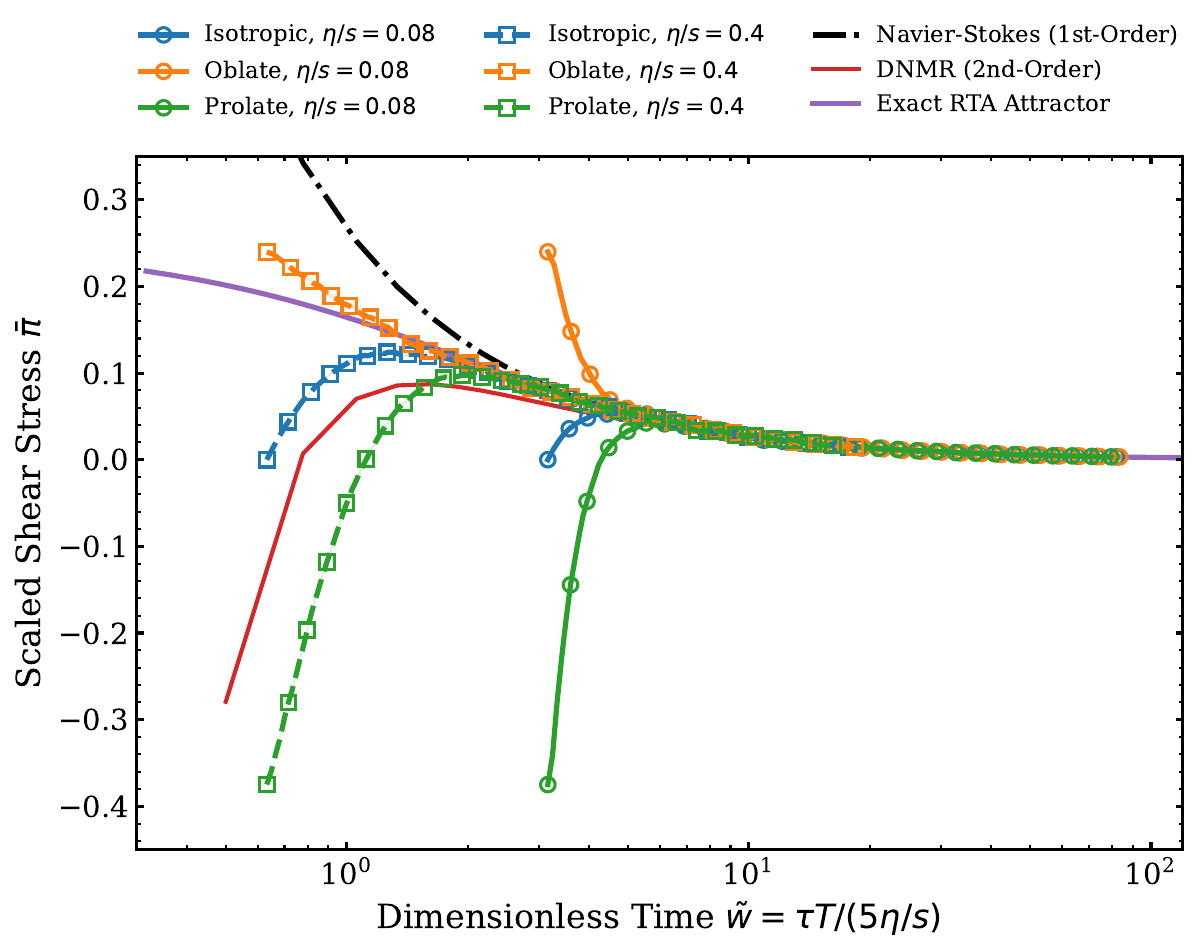}
    \caption{Evolution of the scaled shear stress $\bar{\pi} = [(1/3) \epsilon - T^{zz}]/[(4/3) \epsilon]$ as a function of the dimensionless universal time $\tilde{w} = \tau T/(5\eta/s)$ for a Bjorken expansion. Numerical results generated by the ShARK framework are shown for varying initial momentum anisotropies (isotropic, oblate, and prolate) and transport coefficients ($\eta/s = 0.08$ and $0.4$). These are compared against the first-order Navier-Stokes limit (black dash-dotted line), the second-order DNMR approximation (solid red line), and the exact all-order RTA kinetic attractor (solid purple line).}
    \label{fig:bjorken_attractor}
\end{figure}

In Fig.~\ref{fig:bjorken_attractor}, we present the temporal evolution of $\bar{\pi}$ for six distinct $0+1$D Bjorken flow configurations, varying both the initial momentum-space deformation (isotropic, strongly oblate, and strongly prolate) and the transport coefficients ($\eta/s = 0.08$ and $\eta/s = 0.4$). Although they originate from disparate non-equilibrium states, all six trajectories lose memory of their initial conditions and collapse onto a single, universal non-equilibrium attractor curve at $\tilde{w} \approx 2$.

The resulting kinetic attractor line deviates from the first-order Navier-Stokes limit, $\bar{\pi}_{\rm NS} = 4 / (15\tilde{w})$, and from the second-order DNMR result 
\begin{equation}
    \bar{\pi}_{\rm DNMR} = \frac{4}{15\tilde{w}} - \frac{C}{\tilde{w}^2} + \mathcal{O}\left(\frac{1}{\tilde{w}^3}\right),
\end{equation}
obtained by analytically matching the second-order transport coefficients for a conformal, massless RTA gas, which yields $C = 64/315 \approx 0.203$ \cite{Denicol:2012cn}.

To benchmark ShARK against the theoretical evolution, we numerically solve the integral equation for the RTA Boltzmann energy density. This exact solution was originally formulated in~\cite{Baym:1984np}, generalized for hydrodynamic testing in~\cite{Florkowski:2013lya}, and further explored in~\cite{Strickland:2018ayk, Romatschke:2017vte, Strickland:2019hff, Jaiswal:2019cju}:
\begin{equation}
    \epsilon(\tau) = \Lambda_0^4 D(\tau, \tau_0) R(\xi(\tau, \tau_0)) + \int_{\tau_0}^\tau \frac{d\tau^\prime}{\tau_R} D(\tau, \tau^\prime) \epsilon(\tau^\prime) R(\xi(\tau, \tau^\prime)),
    \label{eqn:volterra_bjorken}
\end{equation}
where $D(\tau_2, \tau_1) = \exp\left(-\int_{\tau_1}^{\tau_2} d\tau^{\prime\prime} \tau_R^{-1}(\tau^{\prime\prime})\right)$ is the exponential damping function and $R(z)$ is the kernel describing the free-streaming momentum anisotropy. Here, the momentum anisotropy parameter driven by the longitudinal Bjorken expansion is defined as $\xi(\tau, \tau^\prime) \equiv (\tau/\tau^\prime)^2 - 1$. To appropriately capture the boundary condition, the initial free-streaming contribution is evaluated at $\xi(\tau, \tau_0) \equiv (1+\xi_0)(\tau/\tau_0)^2 - 1$, thereby propagating the initial momentum-space deformation $\xi_0$ (cf.~the analogous conformal anisotropy construction $\xi_{\rm FS}(\rho; \rho^\prime)$ defined for the Gubser flow in Sec.~\ref{subsec:gubser_flow}). We initialized this reference solution with initial momentum anisotropy ($\xi_0 = 100$) at $\tau_0 \ll 1$, representing the natural free-streaming limit, and ensuring the solution traces the pure universal attractor from its origin. We evaluate the integral memory kernels using a semi-analytic exponential integrator to avoid numerical stiffness and floating-point underflow as the system approaches late-time equilibrium.

As shown in Fig.~\ref{fig:bjorken_attractor}, the ShARK numerical data coincides with the RTA integral solution across the entire evolution. This close agreement across the full evolution, from the far-from-equilibrium regime through the Navier-Stokes asymptote, demonstrates that the stochastic resampling algorithm reproduces the kinetic solution, consistent with a full resummation of hydrodynamic gradient corrections, and validates the framework's capability to evolve the system through regimes where truncated fluid dynamics breaks down.

\subsection{Gubser Flow}
\label{subsec:gubser_flow}

While the 0D homogeneous box technique for the Bjorken expansion validates the longitudinal relaxation kernel, realistic heavy-ion collisions are dominated by rapid transverse expansion. As a first validation of the transverse and radial gradients captured by the model, we turn to the Gubser flow, which describes an azimuthally symmetric fluid. The flow combines simultaneous radial and boost-invariant longitudinal expansion, and is characterized by the conformal symmetry group $SO(3)_q \otimes SO(1,1) \otimes Z_2$~\cite{Gubser:2010ze, Gubser:2010ui}.

In standard Minkowski coordinates, the Gubser flow exhibits large spatial gradients and relativistic velocities that approach the speed of light at large radii, making it notoriously difficult for Eulerian schemes to simulate accurately. However, because the system is composed of a massless conformal gas, the physics is invariant under Weyl rescalings of the metric. It is possible to map the Minkowski Milne metric into a curved de Sitter space $dS_3 \otimes \mathbb{R}$ by performing a coordinate transformation paired with a Weyl rescaling $d\hat{s}^2 = ds^2/\tau^2$~\cite{Gubser:2010ze, Gubser:2010ui, Denicol:2014xca, Denicol:2014tha}. In this curved geometry, the macroscopic fluid four-velocity is static, $\hat{u}^\mu = (1, 0, 0, 0)$, and all macroscopic variables depend exclusively on the de Sitter time $\rho$.

Within the ShARK framework, we adapt the 0D homogeneous box technique to simulate a system macroscopically at rest in de Sitter space. To account for the curvature of $dS_3$, the spatial advection operator is bypassed, and the metric dilution is executed geometrically in momentum space by redshifting the transverse particle momenta at each discrete time step $\Delta\rho$ according to the scale factor
\begin{equation}
    p_\perp(\rho + \Delta\rho) = p_\perp(\rho) \frac{\cosh\rho}{\cosh(\rho + \Delta\rho)}.
\end{equation}

For a given discrete de Sitter time step $\rho$, we evaluate the raw simulated energy density $\epsilon_{\text{sim}}$ directly from the redshifted particle momenta within each cell. However, because the physical cell volume remains fixed while the de Sitter metric expands, the true de Sitter energy density $\hat{\epsilon}$ must manually account for the geometric volume dilution:
\begin{equation}
    \hat{\epsilon}(\rho) = \frac{\epsilon_{\text{sim}}}{\left( \frac{\cosh\rho}{\cosh\rho_0} \right)^2}.
\end{equation}

This corrected energy density defines the effective local temperature via the conformal equation of state, $\hat{\epsilon} = \frac{3}{\pi^2}\hat{T}_{\text{cell}}^4$. Finally, the global fluid temperature $\hat{T}(\rho)$ is obtained from an arithmetic mean across the ensemble of independent cells.

We calculate the scaled shear stress for the entire system by aggregating the energy density $\epsilon$ and the longitudinal component of the stress-energy tensor $T^{zz}$ across all active fluid cells. To remain consistent with the canonical literature for each respective geometry, we adopt the same Gubser sign convention in \cite{Denicol:2014tha}, which defines the shear stress via the longitudinal projection $\pi^\varsigma_\varsigma = P_L - P$. This is opposite to the transverse projection $\Phi = P - P_L$ used in the Bjorken analysis, resulting in an overall sign flip. The global scaled shear stress is therefore extracted as:
\begin{equation}
    \bar{\pi}(\rho) = \frac{T^{zz} - \frac{1}{3}\epsilon}{\frac{4}{3}\epsilon}.
    \label{eqn:scaled_shear_gubser}
\end{equation}

To construct the initial macroscopic state for the Gubser simulations, we initialize the system within the static de Sitter box framework. The fluid is initialized macroscopically at rest in the de Sitter frame, and particle momenta are generated via the RAMBO kernel. The conformal energy density is populated uniformly across a $10 \times 10 \times 10$ Cartesian grid. Rather than using an arbitrary energy density amplitude, the simulation analytically anchors the initial conformal energy density via $\hat{\epsilon}_0 = \cosh^{-8/3}(\rho_0)$ to guarantee that the baseline conformal temperature evaluates to $\hat{T} = 1.0$ when the system crosses $\rho = 0$. We integrate the system using a grid spacing of $0.5$~fm and a conformal time step of $\Delta\rho = 0.001$. Because spatial advection is bypassed in the de Sitter box, the physical grid spacing defines the computational cell volume for the initial discrete particle multiplicity sampling.

The choice of the initial conformal time $\rho_0$ and the corresponding numerical resolution parameters are physically motivated by the specific dynamic features being evaluated:
\begin{itemize}
    \item \textbf{Temperature Evolution and Universal Attractor ($\rho_0 = 0.0$):} To match the standard theoretical integration bounds for the kinetic and hydrodynamic baselines, we initialize these simulations at the point of maximum expansion rate ($\rho_0 = 0.0$). We integrate the temporal evolution up to $\rho = 10.0$ with a particle oversampling factor of $\alpha = 1000$.

    \item \textbf{Shear Inversion and Spatial Profiles ($\rho_0 = -3.0$):} To capture the complete early-time phase-space deformation driven by the longitudinal expansion, these simulations are initialized at $\rho_0 = -3.0$. This negative-time initialization ensures that the artificial isotropic starting conditions ($\bar{\pi}_0 = 0$) fully relax before projecting the spatial gradients onto Minkowski spacetime. Because the initial conformal energy density is anchored via $\hat{\epsilon}_0 = \cosh^{-8/3}(\rho_0)$, the discrete particle multiplicity---which scales with the conformal entropy density as $\cosh^{-2}(\rho_0)$---is exponentially dilute at large negative conformal times. To prevent the computational cells from dropping below the $N < 2$ ballistic threshold and forcing artificial free-streaming prior to hydrodynamization, this configuration uses a larger oversampling factor of $\alpha = 200000$. 
\end{itemize}

Following Denicol et al.~\cite{Denicol:2014tha, Denicol:2014xca}, the ShARK numerical results are compared against five analytical and semi-analytical predictions:
\begin{itemize}
    \item \textbf{Exact Kinetic Theory:} The true analytical solution obtained in~\cite{Denicol:2014tha, Denicol:2014xca}. The temperature $\hat{T}(\rho)$ and shear stress $\bar{\pi}(\rho)$ are found by integrating the energy and shear kernels, $\mathcal{H}(x)$ and $\mathcal{A}(x)$ defined in~\cite{Denicol:2014tha}.
    \item \textbf{Ideal Hydrodynamics:} The zero-viscosity limit follows the closed-form relation $\hat{T}_{\text{ideal}}(\rho) = \hat{T}_0 \cosh^{-2/3}(\rho)$~\cite{Denicol:2014tha}.
    \item \textbf{Free Streaming:} The collisionless limit ($\eta/s \to \infty$), which evaluates the unattenuated initial distribution using the relation $\hat{T}_{\text{FS}}(\rho) = \hat{T}_0 \mathcal{H}^{1/4}\left(\frac{\cosh\rho_0}{\cosh\rho}\right)$~\cite{Denicol:2014tha}.
    \item \textbf{Navier-Stokes (NS):} The first-order limit evaluates the constitutive relation $\bar{\pi}_{\text{NS}} = \frac{4}{3} \frac{\eta/s}{\hat{T}} \tanh\rho$ coupled dynamically to the viscously heated temperature evolution $d\hat{T}/d\rho = -\frac{1}{3} \hat{T} \tanh\rho (2 - \bar{\pi})$~\cite{Denicol:2014tha, Chattopadhyay:2018apf}.
    \item \textbf{Israel-Stewart (IS) and DNMR Hydrodynamics:} We evaluate the second-order approximations by numerically integrating the associated coupled ordinary differential equations obtained in~\cite{Denicol:2014tha} using the Radau method. The IS formulation solves $d\bar{\pi}/d\rho = -\bar{\pi}/\tau_R + \frac{4}{15}\tanh\rho - \frac{4}{3}\bar{\pi}^2 \tanh\rho$~\cite{Denicol:2014tha}. The DNMR formulation incorporates an additional viscous coupling, appending the term $\frac{10}{21}\bar{\pi} \tanh\rho$ to the shear derivative.
\end{itemize}

\subsubsection{Temperature and Shear Stress Evolution}

\begin{figure}[tpb]
    \centering
    \includegraphics[width=0.99\linewidth]{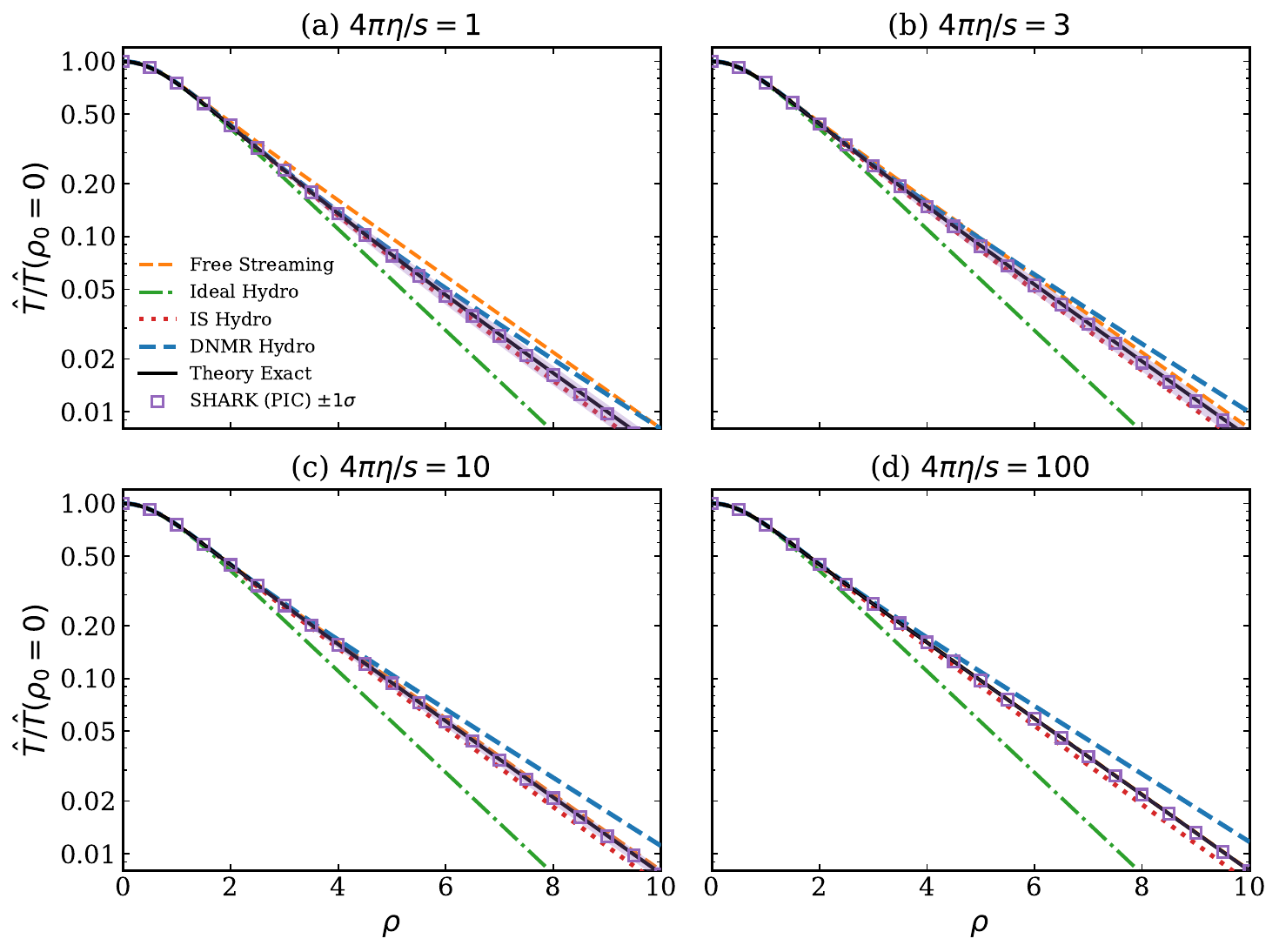}
    \caption{Evolution of the normalized conformal temperature $\hat{T}/\hat{T}(\rho_0=0)$ as a function of de Sitter time $\rho$ for the Gubser expansion. The panels present the cooling trajectories for four specific shear viscosities: (a) $4\pi\eta/s = 1$, (b) $4\pi\eta/s = 3$, (c) $4\pi\eta/s = 10$, and (d) $4\pi\eta/s = 100$. The numerical results from the ShARK solver (purple circles) are shown with $\pm 1\sigma$ statistical uncertainty bands and are compared against the Exact Kinetic Theory (solid black) \cite{Denicol:2014tha, Denicol:2014xca}, Ideal Hydrodynamics (dash-dotted green), Israel-Stewart (IS Hydro, dotted red) and DNMR Hydro (dashed blue).}
    \label{fig:gubser-temperature-evo}
\end{figure}

We begin by evaluating the fireball's thermodynamic cooling. Figure~\ref{fig:gubser-temperature-evo} illustrates the evolution of the normalized temperature for four discrete specific shear viscosities: $4\pi\eta/s = 1, 3, 10, \text{ and } 100$.

 The ideal fluid exhibits the fastest cooling rate. The introduction of finite shear viscosity generates entropy, performing work against the rapid expansion, slowing down the cooling rate, and causing the temperature trajectories to deflect upwards from the ideal limit. The ShARK numerical data reproduces the exact kinetic theory curve within statistical uncertainties across all regimes. In contrast, the macroscopic truncations exhibit noticeable deviations from the exact solution in opposite directions at large de Sitter times: the IS approximation overestimates the cooling rate (falling below the exact curve), while the DNMR formulation underestimates it (diverging above the exact curve).

Beyond the temperature evolution, the Gubser flow exhibits a distinct signature: an inversion in the shear stress~\cite{Denicol:2014tha, Denicol:2014xca, Chattopadhyay:2018apf}. At early de Sitter times ($\rho < 0$), the boost-invariant longitudinal expansion dominates, driving the longitudinal pressure significantly below the transverse pressure. This forces the momentum distribution into an oblate configuration, manifesting as a negative scaled shear stress ($\bar{\pi} < 0$). As the system crosses the origin ($\rho = 0$), the radial transverse expansion overtakes the longitudinal expansion. The pressure anisotropy inverts, driving the distribution into a prolate state ($\bar{\pi} > 0$) that eventually asymptotes toward the collisionless free-streaming limit. 

\begin{figure}[tpb]
    \centering
    \includegraphics[width=0.99\linewidth]{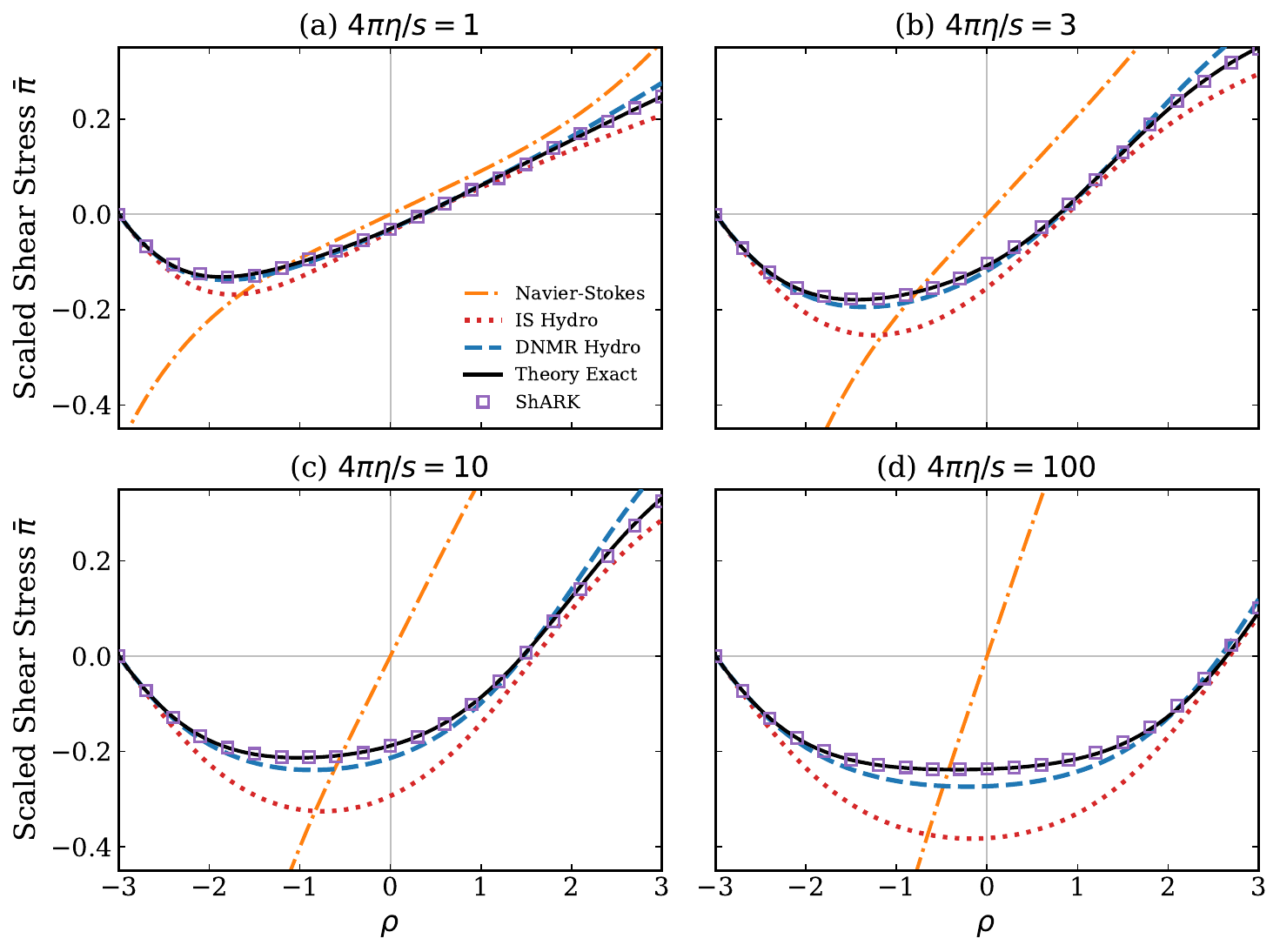}
    \caption{Temporal evolution of the scaled shear stress $\bar{\pi}$ as a function of de Sitter time $\rho$ for the Gubser expansion, initialized at $\rho_0 = -3.0$.
    The panels display the non-equilibrium dynamics for four specific shear viscosities: (a) $4\pi\eta/s = 1$, (b) $4\pi\eta/s = 3$, (c) $4\pi\eta/s = 10$, and (d) $4\pi\eta/s = 100$. The ShARK numerical results (purple circles) are plotted alongside results from Exact Kinetic Theory (solid black), the Navier-Stokes limit (dash-dotted orange) and the second-order approximations, IS Hydro (dotted red) and DNMR Hydro (dashed blue)~\cite{Denicol:2014xca, Chattopadhyay:2018apf}.
    }
    \label{fig:gubser-shear-inv}
\end{figure}

We now extract the temporal evolution of the scaled shear stress $\bar{\pi}(\rho)$ from the ShARK Gubser simulations. The simulation data presented in Fig.~\ref{fig:gubser-shear-inv} were generated by initializing the system at $\rho_0 = -3.0$ and evolving it to $\rho = 3.0$. The ShARK numerical data is benchmarked against four distinct theoretical continuous limits, equally anchored at $\rho_0 = -3.0$.

As shown in Fig.~\ref{fig:gubser-shear-inv}, while ShARK recovers the exact kinetic results, the approximations (IS and DNMR) struggle to capture the shear inversion, deviating at both large negative and positive $\rho$ values~\cite{Denicol:2014xca, Chattopadhyay:2018apf}. Having established this agreement in the temporal domain, we next examine whether it persists in physical space.

\subsubsection{Minkowski Radial Profiles and Spatial Gradients}
\label{subsubsec:gubser_radial_profiles}

We now project the solution back to the physical Minkowski spacetime to evaluate the spatial distribution of the fluid's temperature and shear stress. 

The mapping between the Minkowski Milne coordinates $(\tau, r)$ and the de Sitter time $\rho$ is given by the coordinate transformation:
\begin{equation}
    \rho(\tau, r) = -\text{arcsinh}\left( \frac{1 - q^2\tau^2 + q^2 r^2}{2q\tau} \right),
\end{equation}
where the transverse scale parameter is set to $q = 1.0 \text{ fm}^{-1}$. Using this relation, the macroscopic variables computed in the 0D de Sitter box can be mapped onto a spatial profile at fixed proper time $\tau$. 

The extraction of the simulation data mirrors that used in the temporal evolution analysis. For a given de Sitter state, we compute the conformal energy density $\hat{\epsilon}$ and the scaled shear stress $\bar{\pi}$. The physical Minkowski temperature profile $T(\tau, r)$ is then obtained by undoing the Weyl rescaling:
\begin{equation}
    T(\tau, r) = \frac{\hat{T}(\rho(\tau, r))}{\tau} = \frac{\hat{\epsilon}^{1/4}(\rho(\tau, r))}{\tau}.
\end{equation}
For clarity, we normalize the temperature by the central temperature at $\tau = 1.0 \text{ fm}/c$ ($r=0$, $\rho=0$), denoted as $T_0$. 

It is important to clarify how radial profiles can be extracted out to arbitrary radii (e.g., $r = 10$~fm) despite the simulation employing a finite $10 \times 10 \times 10$ Cartesian grid with a spacing of $0.5$~fm. Because the Gubser benchmark utilizes the 0D homogeneous box technique, it bypasses the spatial advection operator, and the grid does not represent discrete physical coordinates. Instead, the 1000 computational cells act as an ensemble of independent statistical realizations evolving simultaneously in conformal time $\rho$. The physical grid spacing defines only the computational cell volume $V_{\rm cell}$, which, combined with the oversampling parameter $\alpha$, dictates the local test-particle occupancy required to manage Monte Carlo Poisson noise. Consequently, the mapping to Minkowski space is temporal rather than spatial. For any arbitrary physical radius $r$ at a fixed proper time $\tau$, the analytical coordinate transformation determines the corresponding conformal time $\rho(\tau, r)$. The framework simply retrieves the globally averaged thermodynamic state of the 0D ensemble at that specific $\rho$ and applies the $1/\tau$ Weyl rescaling. Provided the simulation is initialized sufficiently deep in the de Sitter past (large negative $\rho_0$) to encompass the required evaluation times, the radial profile can be extended indefinitely without encountering finite-size boundary artifacts.

We plot the magnitude of the dominant shear stress component $\bar{\pi}^\xi_\xi$, which corresponds to $4|\bar{\pi}|$ in the standard convention~\cite{Denicol:2014tha}. The theoretical baselines for this spatial evaluation, including exact kinetic theory, the Israel-Stewart (IS) approximation, and ideal hydrodynamics, are generated utilizing the continuous limits established in Sec.~\ref{subsec:gubser_flow}. All spatial profiles mask out regions lying outside the causal initialization boundary ($\rho < \rho_0$).

\begin{figure}
    \centering
    \includegraphics[width=0.99\linewidth]{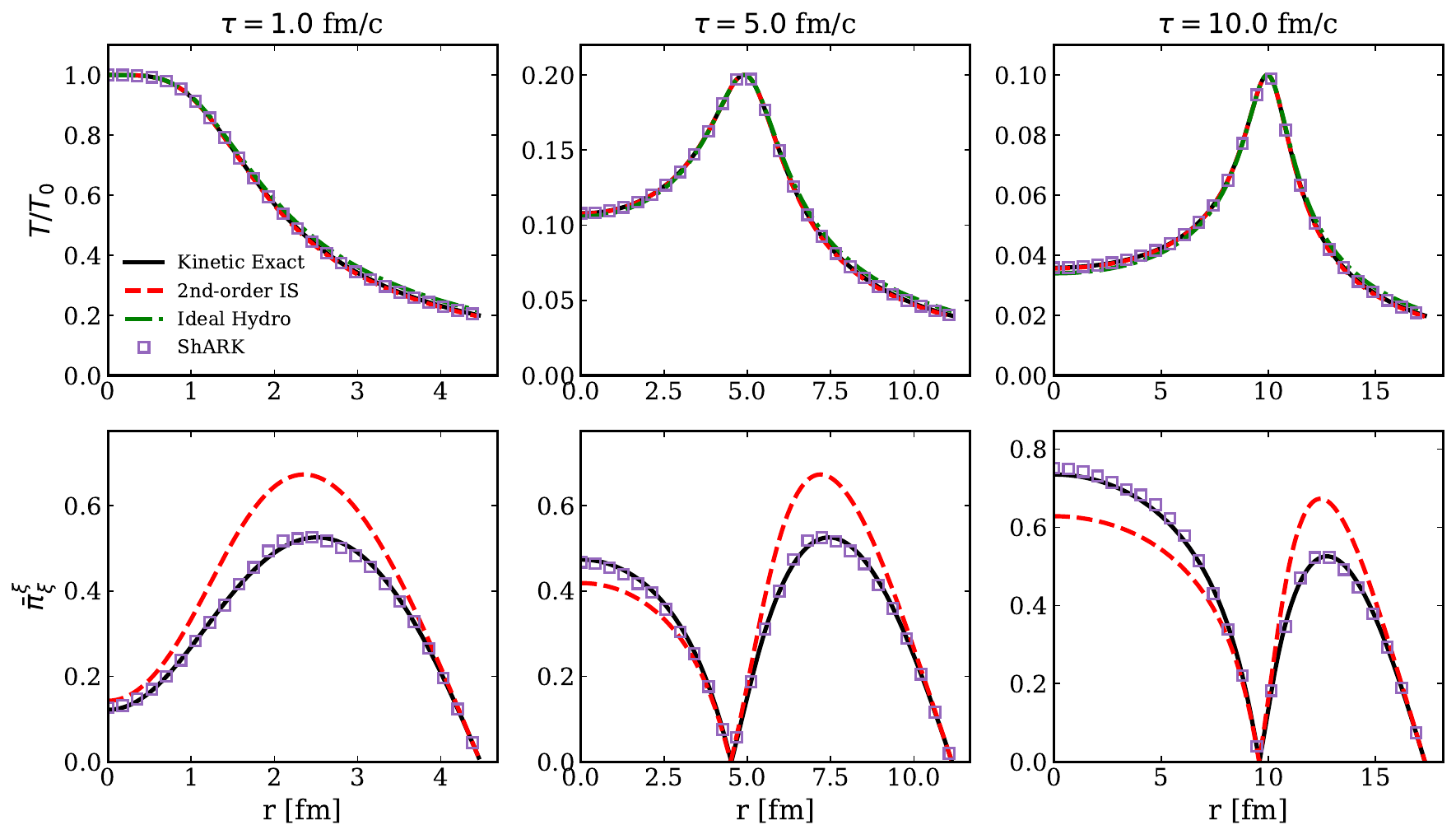}
    \caption{Minkowski radial profiles of the normalized temperature $T/T_0$ (top row) and the shear stress magnitude $\bar{\pi}^\xi_\xi$ (bottom row) for the Gubser expansion at proper times $\tau = 1.0, 5.0$, and $10.0$ fm/$c$. The specific shear viscosity is set to the near-ideal regime, $4\pi\eta/s = 1$. The ShARK simulation (purple circles) is compared to Exact Kinetic Theory (solid black), Ideal Hydrodynamics (dash-dotted green) and the second-order Israel-Stewart approximation (dashed red)~\cite{Denicol:2014tha, Denicol:2014xca}.}
    \label{fig:gubser-profile-4pietas1}
\end{figure}

\begin{figure}
    \centering
    \includegraphics[width=0.99\linewidth]{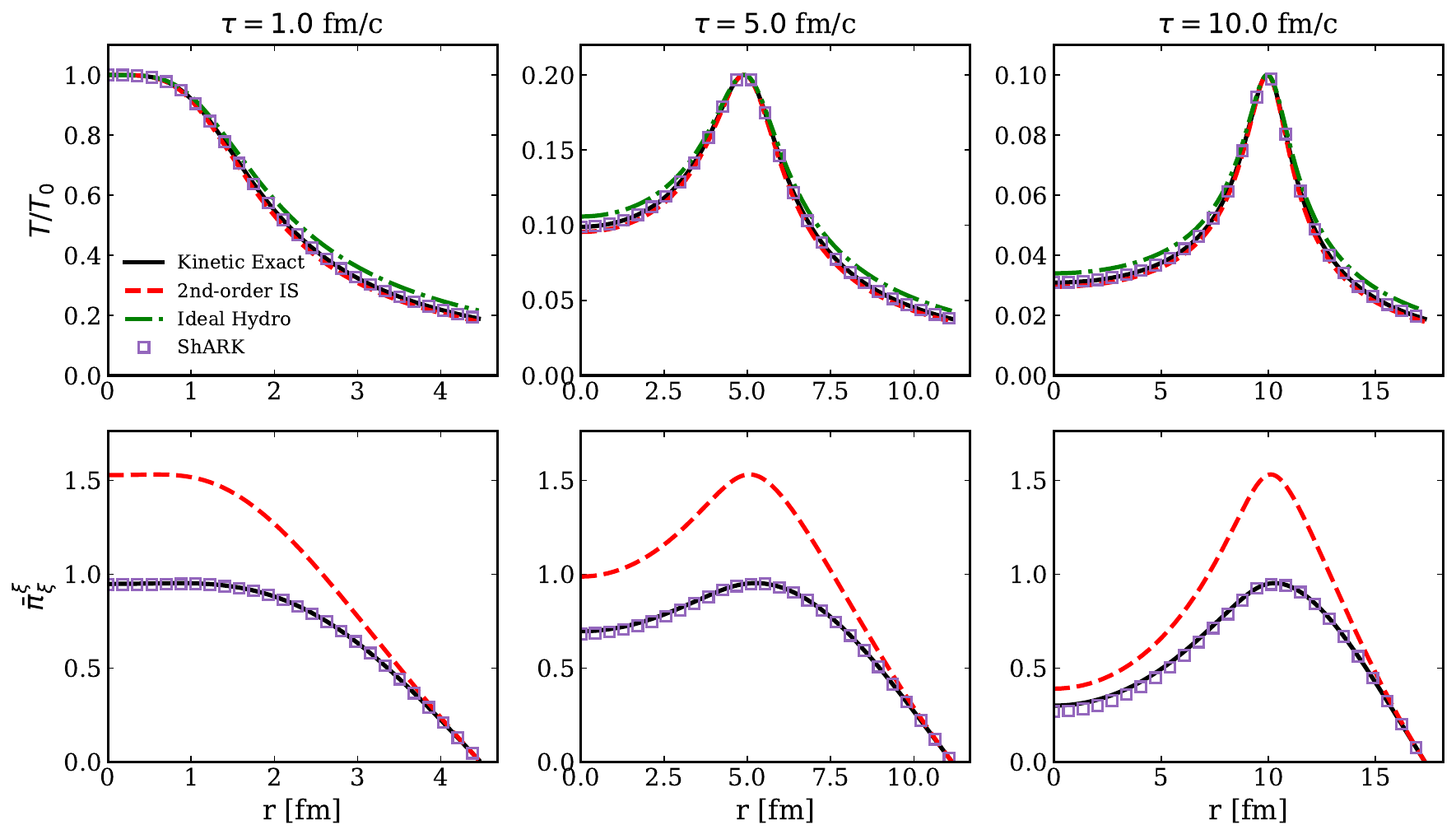}
    \caption{Minkowski radial profiles of the normalized temperature $T/T_0$ (top row) and the shear stress magnitude $\bar{\pi}^\xi_\xi$ (bottom row) for the Gubser expansion at proper times $\tau = 1.0, 5.0$, and $10.0$ fm/$c$. The specific shear viscosity is set to the highly viscous regime, $4\pi\eta/s = 100$. The ShARK simulation (purple circles) is compared to Exact Kinetic Theory (solid black), Ideal Hydrodynamics (dash-dotted green) and the second-order Israel-Stewart approximation (dashed red)~\cite{Denicol:2014tha, Denicol:2014xca}.}
    \label{fig:gubser-profile-4pietas100}
\end{figure}

Figures~\ref{fig:gubser-profile-4pietas1} and \ref{fig:gubser-profile-4pietas100} present the radial profiles of the normalized temperature $T/T_0$ (top rows) and the scaled shear stress component $\bar{\pi}^\xi_\xi$ (bottom rows) for specific shear viscosities $4\pi\eta/s = 1$ and $100$, respectively. The profiles are shown as snapshots at three proper times: $\tau = 1.0, 5.0, \text{ and } 10.0 \text{ fm}/c$. 

At early times ($\tau = 1.0 \text{ fm}/c$), the temperature profile resembles a Gaussian centered at $r=0$. As the system evolves to $\tau = 5.0$ and $10.0 \text{ fm}/c$, the radial expansion pushes the bulk of the energy density outward, and the peak temperature propagates to larger radii. Simultaneously, the shear stress develops a sharp dip to zero at the radius where the expansion rates balance and the pressure anisotropy inverts.

The ShARK results accurately trace the exact kinetic baselines across all proper times, radii, and viscosities. The stochastic solver accurately captures both the expanding thermal profile and the exact shear stress evolution across both the smaller ($4\pi\eta/s = 1$) and larger ($4\pi\eta/s = 100$) viscosity regimes. The second-order Israel-Stewart approximation struggles consistently with the dissipative amplitudes. Even at the smaller viscosity ($4\pi\eta/s = 1$), the truncation visibly deviates from the exact shear-stress magnitude. This mismatch is magnified in the far-from-equilibrium regime ($4\pi\eta/s = 100$), providing evidence that the stochastic formulation correctly resums the higher-order spatial gradients that truncated fluid dynamics cannot resolve.

\subsubsection{The Universal Non-Equilibrium Attractor}
\label{subsubsec:gubser_attractor}

Having established that ShARK reproduces the non-equilibrium dynamics for a fixed initial condition, both temporally and spatially, a natural question is whether this agreement holds across initial conditions. We now verify the fluid's ability to lose memory of its initial microscopic configuration within the Gubser expansion by evaluating the convergence of widely divergent initial states onto a universal non-equilibrium attractor curve. This independence from initial conditions is particularly relevant, since realistic heavy-ion collisions begin from a poorly constrained, far-from-equilibrium initial state.

To ensure that the ShARK simulation shares an identical thermodynamic history with the theoretical predictions, rather than a "cold start" at negative $\rho$, the system is initialized at the point of maximum expansion rate, $\rho_0 = 0.0$. We extract the scaled shear stress continuously from the 0D de Sitter box using the relation in Eq.~\eqref{eqn:scaled_shear_gubser}, which dynamically tracks the fluid's deformation as it evolves to $\rho = 10.0$.

Evaluating the analytical solution of the RTA Boltzmann equation from an initialization at $\rho_0 = 0$ is numerically challenging due to branch-cut singularities in the standard fixed-coordinate kernels~\cite{Denicol:2014tha}. To avoid this, we evaluate the exact solution using the Romatschke-Strickland (RS) anisotropic formulation~\cite{Romatschke:2003ms, Behtash:2017wqg, Chattopadhyay:2018apf}. In this framework, the dynamics are mapped directly to a geometric free-streaming momentum anisotropy parameter: 
\begin{equation}
    \xi_{FS}(\rho; \rho') = -1 + \left(\frac{\cosh\rho'}{\cosh\rho}\right)^2 \ .
\end{equation}
Using $\xi_{FS}$, the temperature $\hat{T}(\rho)$ and shear stress $\bar{\pi}(\rho)$ are extracted by iteratively solving the integral equations with the RS thermodynamic moments, $\mathcal{R}_{200}(\xi)$ and $\mathcal{R}_{220}(\xi)$~\cite{Behtash:2017wqg, Chattopadhyay:2018apf}. These moments remain real and well-behaved across the entire conformal domain, ensuring a stable evaluation. 

The kinetic attractor is supplemented by two fluid dynamic limits, both anchored at $\hat{T}_0 = 1.0$ and $\rho_0 = 0.0$. We evaluate the Navier-Stokes (NS) and second-order Israel-Stewart (IS) approximations using the respective constitutive relations and coupled differential equations established in Section~\ref{subsec:gubser_flow}~\cite{Denicol:2014tha, Behtash:2017wqg, Chattopadhyay:2018apf}.

\begin{figure}
    \centering
    \includegraphics[width=0.99\linewidth]{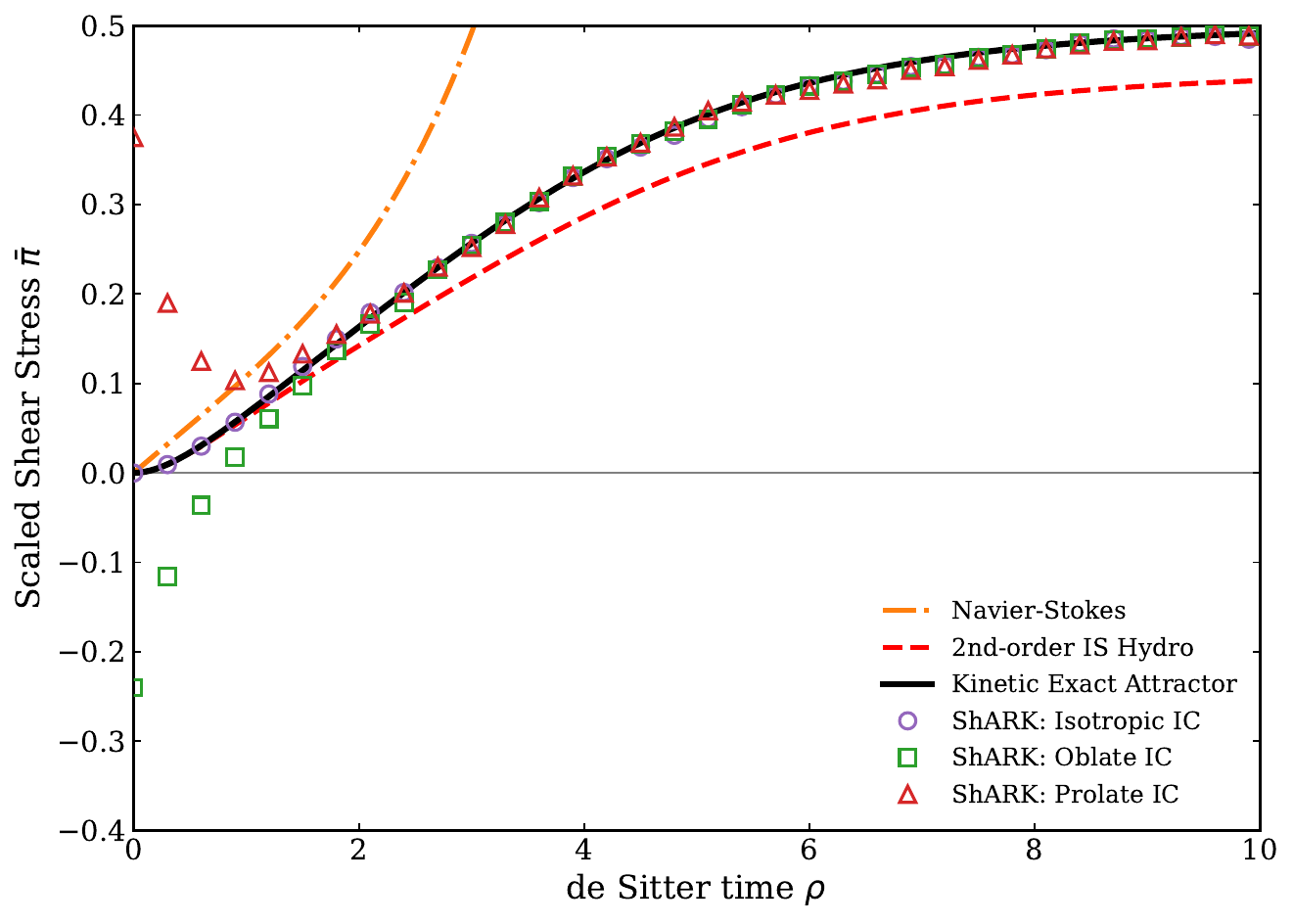}
    \caption{Evolution of the scaled shear stress $\bar{\pi}$ as a function of de Sitter time $\rho$ demonstrating the universal non-equilibrium attractor of the Gubser flow for a specific shear viscosity of $4\pi\eta/s = 1$. The ShARK simulation is initialized in three distinct microscopic states: an isotropic thermal equilibrium ($\bar{\pi}_0 = 0$, purple circles), an oblate deformation ($\bar{\pi}_0 < 0$, green squares), and a prolate deformation ($\bar{\pi}_0 > 0$, red triangles). The numerical data is compared to the Kinetic Exact Attractor (solid black line), the first-order Navier-Stokes (dash-dotted orange) and second-order IS Hydro (dashed red) \cite{Denicol:2014tha, Behtash:2017wqg, Chattopadhyay:2018apf}.}
    \label{fig:gubser-attr}
\end{figure}

Figure~\ref{fig:gubser-attr} presents the resulting phase-space evolution for a specific shear viscosity of $4\pi\eta/s = 1$. The ShARK simulation is executed for three distinct initial states: an isotropic thermal equilibrium state ($\bar{\pi}_0 = 0$), an oblate deformation ($\bar{\pi}_0 < 0$), and a prolate deformation ($\bar{\pi}_0 > 0$). 

Despite originating from different microscopic configurations, all three simulated trajectories lose memory of their initialization within $\Delta\rho \approx 2$, collapsing onto a single unified trajectory that traces the solid black Exact Kinetic Attractor line. This overlap is consistent with a full resummation of the infinite hierarchy of transport moments, matching an exact, all-order solution to the relativistic Boltzmann equation.

\subsection{Realistic Initial Geometries and Numerical Convergence}
\label{subsec:optical_glauber}

While analytical benchmarks such as the Bjorken and Gubser flows validate the core evolution equations under idealized symmetries, phenomenological simulations require robust performance against realistic nuclear geometries. To evaluate the ShARK solver under these conditions, we perform a self-convergence validation using 3D optical Glauber initial conditions. This section presents a numerical convergence analysis across spatial resolution ($\Delta x$), temporal step size ($\Delta t$), and particle oversampling statistics ($\alpha$).

We note that this convergence analysis is intended as a self-consistency check of the numerical scheme under a representative, realistic initial condition, rather than an exhaustive convergence study across all observables, proper times, and collision geometries relevant for phenomenological applications. In particular, the extracted convergence order can depend on the proper time $\tau$ at which it is evaluated, reflecting the interplay between the local smoothness of the evolving profile and the numerical scheme's local order reduction near steep gradients, a well-known property of TVD/flux-limited schemes~\cite{VanLeer1977, Harten1983}. A comprehensive convergence study assessing observable-, time-, and geometry-dependence is deferred to future work.

\subsubsection{Initial Condition Construction}

We construct the initial energy density profile by combining a transverse nuclear overlap with a longitudinal spacetime-rapidity envelope.

For the transverse plane, the nuclear thickness functions $T_A(x,y)$ and $T_B(x,y)$ are computed by integrating the standard Woods-Saxon nuclear density distributions along the longitudinal axis \cite{Miller:2007ri}:
\begin{equation}
    T(x,y) = \int_{-\infty}^{\infty} \frac{\rho_0}{1 + \exp\left(\frac{\sqrt{x^2 + y^2 + z^2} - R}{a}\right)} dz \, ,
\end{equation}
where $\rho_0$ is the nuclear saturation density, $R$ is the nuclear radius, and $a$ is the surface diffuseness parameter. Assuming an optical Glauber framework \cite{Miller:2007ri} with an inelastic nucleon-nucleon cross section $\sigma_{NN}$, we evaluate the local participant density $N_{\text{part}}(x,y)$ at a given impact parameter $b$. The initial transverse energy density $\varepsilon_T(x,y)$ is then scaled proportionally to $N_{\text{part}}(x,y)$.

To embed this 3D structure onto the Cartesian grid, the transverse profile is scaled with a Gaussian spacetime-rapidity envelope $H(\eta_s) = \exp(-\eta_s^2 / 2\sigma_\eta^2)$ and a geometric proper-time dilution factor that accounts for the longitudinal Bjorken expansion \cite{Bjorken:1982qr}:
\begin{equation}
    \varepsilon(x, y, z) = \varepsilon_T(x,y) \exp\left(-\frac{\eta_s^2}{2\sigma_\eta^2}\right) \left(\frac{\tau_0}{\tau_{\text{local}}}\right)^{4/3} \, ,
\end{equation}
where the local proper time is given by $\tau_{\text{local}} = \sqrt{t^2 - z^2}$, and $\eta_s = \text{arctanh}(z/t)$ is the spacetime rapidity evaluated at the initial laboratory time $t = \tau_0$. Finally, the fluid is initialized with a longitudinal boost velocity $v_z = z / t$, corresponding to ideal Bjorken flow, while the initial transverse flow velocities are set to zero ($v_x = v_y = 0$).

The parameters represent a generic high-energy Oxygen-Oxygen (O-O) collision and are listed in Table~\ref{tab:glauber_params}. This provides a rapidly expanding 3D geometry with steep spatial gradients. The Woods-Saxon parameters $R = 2.608$~fm and $a = 0.513$~fm define the $^{16}\text{O}$ nuclear density profile~\cite{Sievert:2019zjr}, normalized against the standard nuclear saturation density of $\rho_0 = 0.16$~fm$^{-3}$. We set the remaining parameters to typical collider energy scales and idealized numerical conditions to benchmark the solver.

\begin{table}[h]
    \centering
    \begin{tabular}{llc}
        \hline\hline
        $b$ & Impact parameter & $0.0$ fm \\
        $\varepsilon_0$ & Central energy density scale & $100.0$ GeV/fm$^3$ \\
        $R$ & Woods-Saxon radius for $^{16}\text{O}$ & $2.608$ fm \\
        $a$ & Woods-Saxon skin thickness & $0.513$ fm \\
        $\rho_0$ & Nuclear saturation density & $0.16$ fm$^{-3}$ \\
        $\sigma_{NN}$ & Inelastic nucleon-nucleon cross section & $6.4$ fm$^2$ \\
        $\sigma_\eta$ & Gaussian width of the longitudinal envelope & $2.0$ \\
        \hline\hline
    \end{tabular}
    \caption{Parameters for the optical Glauber O-O initial condition used in the convergence tests of this section.}
    \label{tab:glauber_params}
\end{table}

Because this section serves as a numerical convergence test, we simulate a central collision ($b=0$) initialized at proper time $\tau_0 = 0.6$~fm/$c$, setting the specific shear viscosity to $\eta/s = 0.08$ across all runs. The central energy density amplitude $\varepsilon_0 = 100.0$~GeV/fm$^3$ is a normalization chosen so that the dense core of the fireball is deeply within the continuum hydrodynamic regime ($N_{\rm cell} \gg 1$) at early times, matching the central scale used for the Bjorken benchmark (Sec.~\ref{subsec:bjorken}). Finally, we model the longitudinal structure of the 3D fireball with a Gaussian rapidity envelope of width $\sigma_\eta = 2.0$, creating a smooth, finite pseudo-rapidity plateau that mitigates the artificial longitudinal edge effects inherent to strict boost invariance.

\subsubsection{Spatial Convergence and Grid Resolution}

We now assess the numerical fidelity of the solver's spatial advection scheme by examining convergence under grid refinement. To establish Cauchy convergence of the spatial advection scheme, we evolve the system across four different Cartesian grid resolutions: $\Delta x \in \{1.0, 0.4, 0.2, 0.1\}$~fm. To isolate the spatial truncation error, we hold the time step fixed at $\Delta t = 0.02$~fm/$c$. The oversampling factor $\alpha$ is scaled inversely with the cell volume $V_{\rm cell} = \Delta x^3$, ensuring that finite-$N$ stochastic fluctuations remain consistent across all resolutions. The simulations utilize $\alpha \in \{16.0, 250.0, 2000.0, 16000.0\}$ for the respective grid spacings.

\begin{figure*}[tpb]
    \centering
    \includegraphics[width=\textwidth]{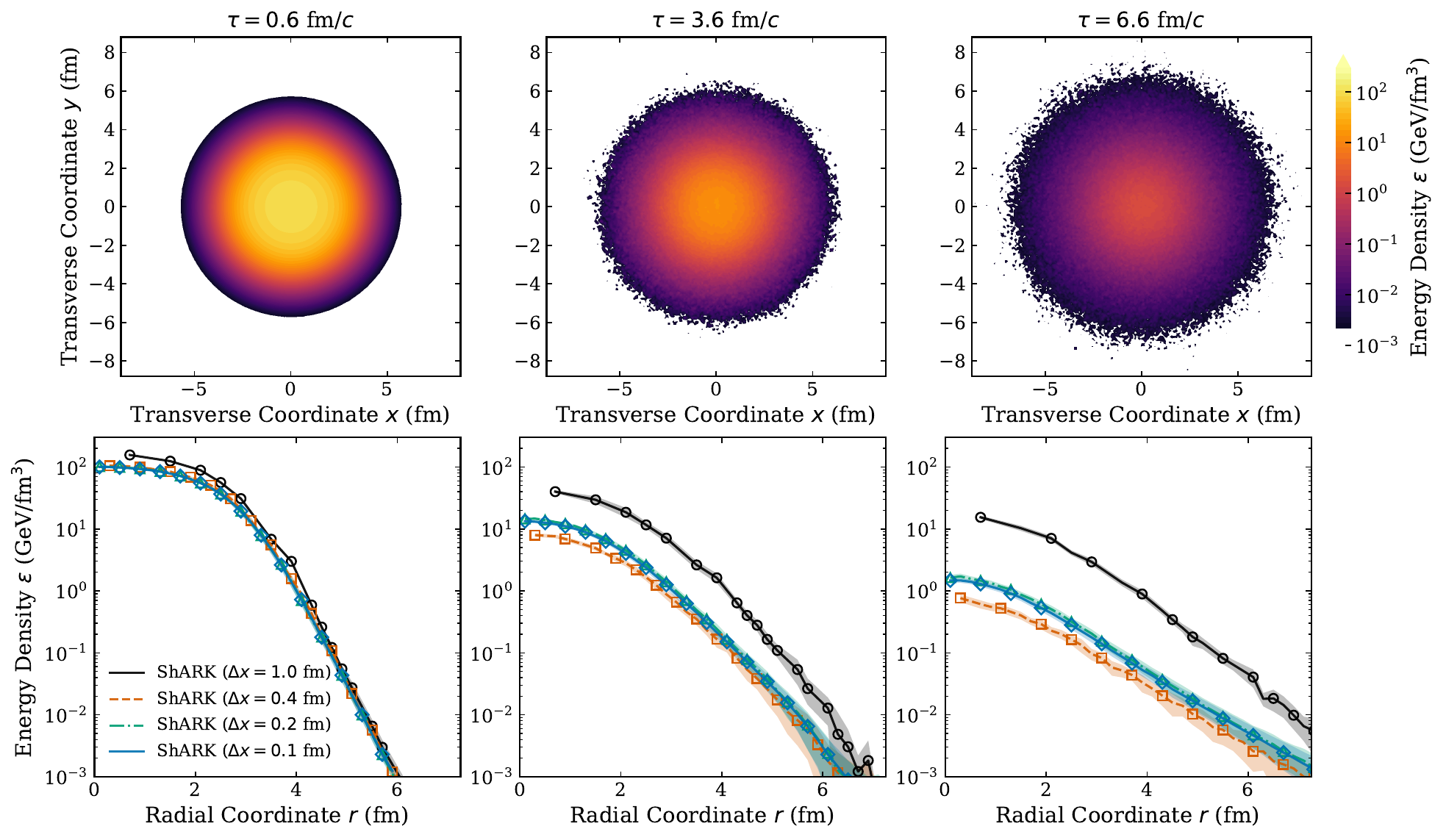}
    \caption{Transverse spatial convergence of ShARK for an optical Glauber initial condition. \textbf{Top row:} 2D heatmaps of the energy density in the transverse $x-y$ midplane ($z=0$) at proper times $\tau = 0.6, 3.6$, and $6.6$ fm/$c$, extracted from the highest resolution grid ($\Delta x = 0.1$ fm). \textbf{Bottom row:} Corresponding 1D radial profiles of the energy density evaluated across four different spatial grid resolutions ($\Delta x = 1.0, 0.4, 0.2$, and $0.1$ fm).}
    \label{fig:transverse_glauber}
\end{figure*}
Figure \ref{fig:transverse_glauber} illustrates the transverse expansion dynamics. The top row displays 2D heatmaps of the energy density in the $x-y$ midplane ($z=0$) extracted from the finest grid ($\Delta x = 0.1$~fm). The solver preserves the rotational symmetry of the central collision as the fluid expands radially into the vacuum. The bottom row presents the radial profiles of the energy density for all four grid spacings. At early times ($\tau = 0.6$~fm/$c$), all resolutions capture the initial Woods-Saxon overlap. However, as radial flow develops, the coarser grids show numerical diffusion, artificially smearing the energy density outward and over-predicting the extent of the dilute corona. In contrast, the $\Delta x = 0.2$~fm and $\Delta x = 0.1$~fm profiles are consistent within statistical uncertainties across all radii and time steps, indicating convergence to the continuum limit.

\begin{figure*}[tpb]
    \centering
    \includegraphics[width=\textwidth]{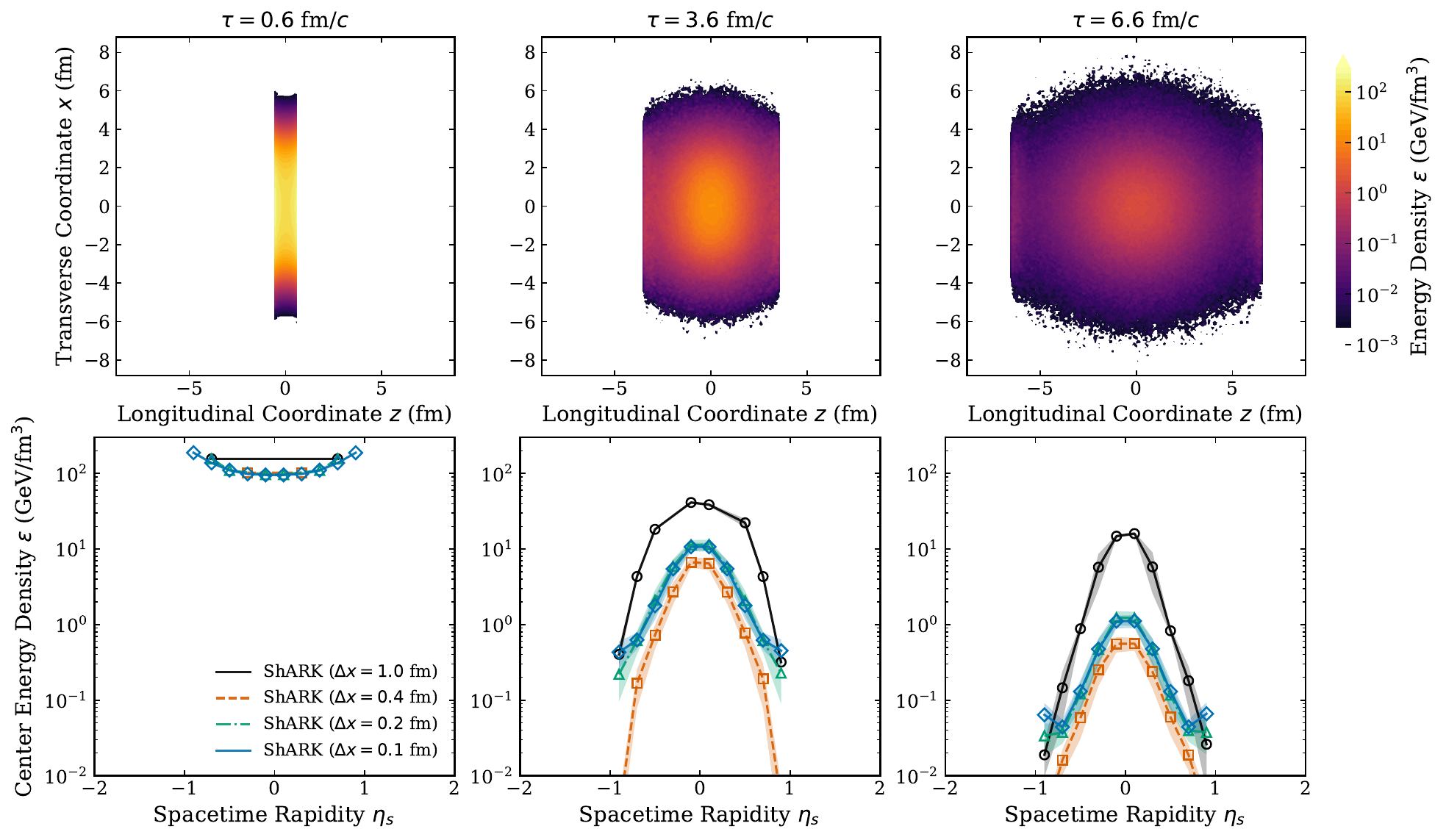}
    \caption{Longitudinal spatial convergence of ShARK for the same optical Glauber initialization. \textbf{Top row:} 2D heatmaps of the energy density in the longitudinal $z-x$ plane ($y=0$) at proper times $\tau = 0.6, 3.6$, and $6.6$ fm/$c$, utilizing the $\Delta x = 0.1$ fm grid. \textbf{Bottom row:} 1D proper energy density profiles evaluated at the transverse center as a function of spacetime rapidity $\eta_s$. }
    \label{fig:longitudinal_glauber}
\end{figure*}
Figure \ref{fig:longitudinal_glauber} presents the corresponding longitudinal evolution. The top row isolates the $z-x$ plane for the highest resolution run. At the initial time $\tau = 0.6$~fm/$c$, the system is confined to the forward lightcone ($|z| < t$), reflecting the Lorentz contraction of the initial geometry. As the system evolves, the advection resolves the transition from a contracted pancake to a fully three-dimensional expanding fluid. The bottom row evaluates the proper energy density as a function of spacetime rapidity $\eta_s$. The initialized Gaussian envelope ($\sigma_\eta = 2.0$) is recovered. Similar to the transverse plane, the longitudinal profiles on the $\Delta x = 0.2$~fm and $0.1$~fm grids agree within statistical uncertainties, indicating that the solver captures the longitudinal dilution ($\varepsilon \propto \tau^{-4/3}$) of the system without significant spurious numerical viscosity along the beamline.

To formalize the qualitative convergence observed in the spatial profiles, we computed the relative area-weighted $L_2$ error of the midplane energy density against the $\Delta x = 0.1$~fm reference grid, following standard grid-refinement evaluation methods~\cite{LeVeque_2002}. Evaluated within the dense hydrodynamic core ($r \leq 2.0$~fm) at late times ($\tau \gtrsim 2.6$~fm/$c$), the extracted convergence order lies consistently between $p \approx 1.7$--$1.8$, modestly below the scheme's nominal second-order design accuracy.

\subsubsection{Temporal Convergence and Collisionless Scaling}

Having established Cauchy convergence in the spatial domain at $\Delta x = 0.2$~fm, we now turn to the temporal resolution. In a particle-in-cell kinetic framework, the temporal step $\Delta t$ determines how accurately the solver resolves both the geometric free-streaming advection and the local scattering probability set by the relaxation-time approximation.

To isolate the temporal truncation error, we lock the spatial grid at the converged resolution ($\Delta x = 0.2$~fm, producing a $24.0$~fm physical box) and maintain a constant test-particle oversampling factor ($\alpha = 2000$, shown in the following subsection to be sufficient to suppress statistical noise). We then evolve the Oxygen-Oxygen system across four temporal resolutions: $\Delta t \in \{0.10, 0.04, 0.02, 0.01\}$~fm/$c$.

\begin{figure*}[tpb]
    \centering
    \includegraphics[width=\textwidth]{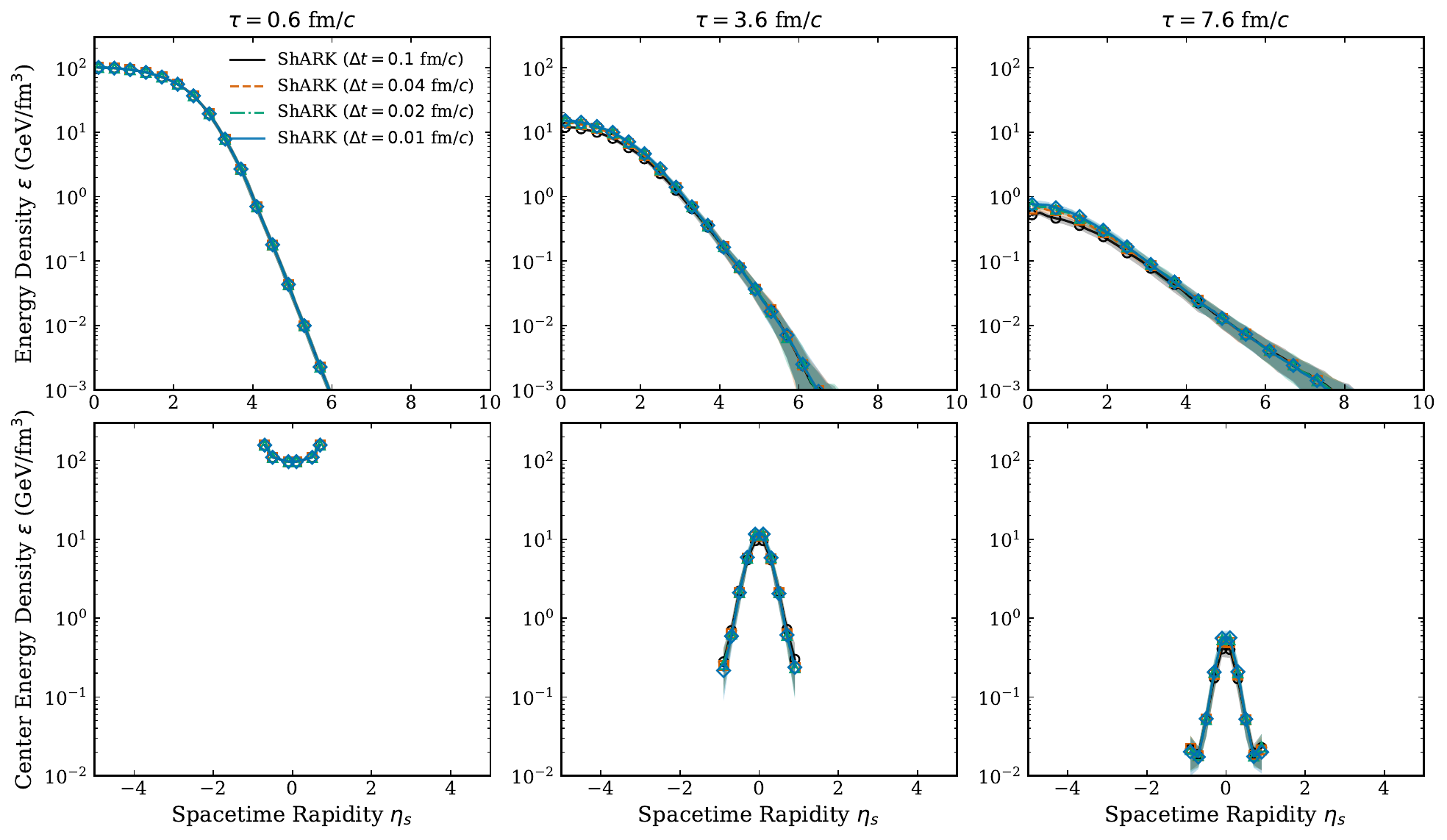}
    \caption{Temporal convergence of the ShARK solver. The top row presents the 1D radial profiles of the energy density at the midplane, while the bottom row displays the longitudinal profiles as a function of spacetime rapidity. The temporal step sizes range from a coarse $\Delta t = 0.10$ fm/$c$ down to a fine $\Delta t = 0.01$ fm/$c$.}
    \label{fig:temporal_glauber}
\end{figure*}

Figure \ref{fig:temporal_glauber} visualizes the temporal convergence through the midplane radial profiles (top row) and longitudinal spacetime rapidity profiles (bottom row) evaluated at $\tau = 0.6, 3.6,$ and $6.6$~fm/$c$. The solver exhibits a distinct bifurcation in convergence behavior depending on the local fluid density.

At large radii ($r > 4.0$~fm) and large rapidities ($|\eta_s| > 1.0$), the energy density drops significantly. In these dilute regions, the effective relaxation time ($\tau_R \propto \frac{\eta/s}{T}$) diverges, and the fluid rapidly transitions into a collisionless free-streaming corona. Because the particle spatial propagation is solved via a linear geometric update ($x = x_0 + v_x \Delta t$), the macroscopic trajectory of non-interacting test particles is exact regardless of the step size. Consequently, the temporal truncation error in the tail is negligible, and all $\Delta t$ evaluations overlap within statistical limits. 

Conversely, at small radii ($r < 2.0$~fm) and mid-rapidity ($\eta_s \approx 0$), the dense interacting core has maximum temperatures and minimal relaxation times. At early proper times, a coarse step such as $\Delta t = 0.10$~fm/$c$ approaches or exceeds the local $\tau_R$, introducing numerical stiffness. This stiffness artificially suppresses early-time scattering rates and miscalculates the corresponding pressure gradients. Integrated over the full hydrodynamic expansion to $\tau = 6.6$~fm/$c$, this early-time error compounds, resulting in the visible deviation of the coarse profiles at the origin.

The profiles extracted from the $\Delta t = 0.02$~fm/$c$ and $\Delta t = 0.01$~fm/$c$ simulations map directly onto each other across the entire spatial and temporal domain. This confirms that a temporal resolution of $\Delta t = 0.02$~fm/$c$ is sufficient for this collision geometry to suppress numerical stiffness in the core and yields a temporally converged solution for realistic heavy-ion geometries.

\subsubsection{Statistical Convergence and Poisson Noise}

Because the ShARK solver utilizes a stochastic PIC methodology, the thermodynamic fields extracted from the particle ensemble are subject to statistical fluctuations. To verify that the simulations represent the true continuous fluid limit, we evaluate the convergence of the energy density profiles against the test-particle oversampling factor, $\alpha$. We lock the grid resolution and time step to their converged values ($\Delta x = 0.2$~fm, $\Delta t = 0.02$~fm/$c$), and sweep $\alpha$ from a highly dilute $\alpha = 250$ to $\alpha = 4000$.

\begin{figure*}[tpb]
    \centering
    \includegraphics[width=\textwidth]{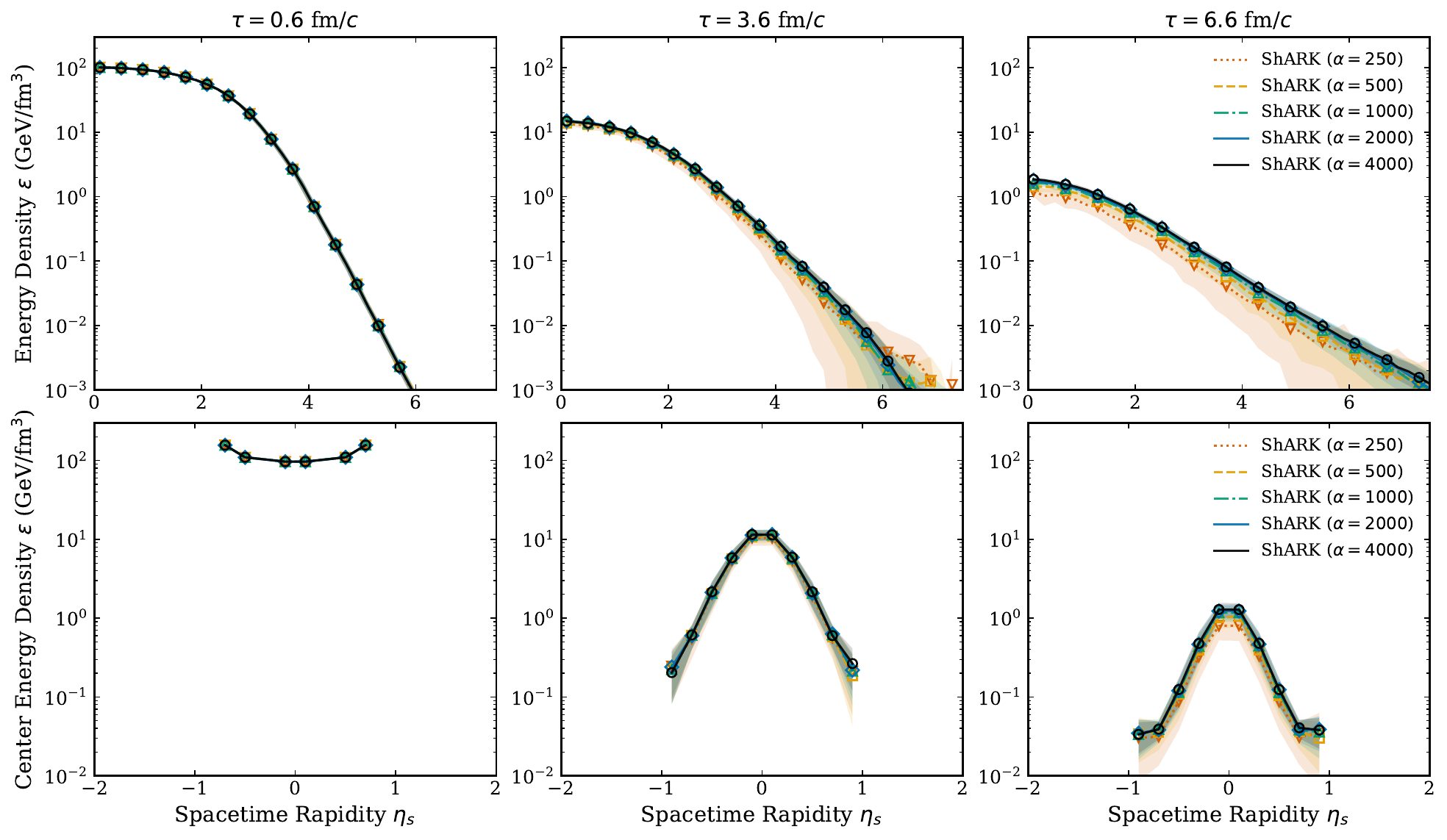}
    \caption{Statistical convergence of the ShARK solver for an optical Glauber O-O initial condition as a function of the test-particle oversampling factor $\alpha$. \textbf{Top row:} 1D radial profiles of the midplane energy density at proper times $\tau = 0.6, 3.6$, and $6.6$ fm/$c$. \textbf{Bottom row:} Corresponding longitudinal energy density profiles evaluated at the transverse center as a function of spacetime rapidity $\eta_s$. The shaded bands represent the standard deviation within the spatial bins.}
    \label{fig:statistical_glauber}
\end{figure*}

Figure \ref{fig:statistical_glauber} illustrates the statistical stabilization of the Oxygen-Oxygen expansion. As expected from Poisson statistics, the relative statistical uncertainty in the energy density, shown by the shaded standard deviation bands, scales as $1/\sqrt{N_{\text{cell}}}$, and steadily shrinks as $\alpha$ increases. At lower oversampling factors ($\alpha = 250, 500$), the dilute corona is dominated by statistical noise, which artificially distorts the pressure gradients and causes the mean energy density profile to sag in the large-radii tails at late times. As oversampling increases to $\alpha = 2000$, the profiles lock into a stable, continuous limit. The overlap between the $\alpha = 2000$ and $\alpha = 4000$ evaluations confirms that $\alpha = 2000$ is sufficient for this collision geometry to resolve the fireball's thermodynamic gradients without introducing unphysical statistical dissipation.

\section{Conclusions}
\label{sec:conclusions}

In this work, we introduced ShARK, a 3D stochastic particle-in-cell transport framework, establishing its theoretical foundation as a first-principles solver for the relativistic Anderson--Witting Boltzmann equation. By analyzing the $N$-body master equation, we showed that a Markovian operator-splitting scheme using the RAMBO algorithm for full-redraw momentum updates natively generates the microcanonical Lorentz-invariant phase space (LIPS). In the thermodynamic limit ($N \gg 1$), this discrete phase-space sampling yields the continuous J\"{u}ttner-Boltzmann equilibrium distribution and recovers the macroscopic relaxation-time approximation.

To validate the numerical implementation, we tested the framework against both analytical benchmarks and 3D self-convergence tests. Using a homogeneous 0D box technique for the Bjorken expansion and a conformal mapping to a static $dS_3 \otimes \mathbb{R}$ geometry for the Gubser flow, we isolated the kinetic dissipation from Cartesian advection artifacts. In both cases, the Monte Carlo framework reproduced the analytical kinetic solutions. Notably, ShARK captured the all-order non-equilibrium attractor and the shear stress inversion of the Gubser flow, correctly resolving far-from-equilibrium dynamics where truncated second-order approximations such as Israel-Stewart and DNMR hydrodynamics break down. Self-convergence tests with fully 3D optical Glauber initial conditions further confirmed the stability and Cauchy convergence of the Eulerian advection scheme, showing that the solver handles the steep spatial gradients characteristic of realistic nuclear collisions without introducing significant spurious numerical diffusion.

A structural advantage of this stochastic formulation is its native handling of the finite-$N$ kinetic regime. Because the microcanonical phase space imposes strict kinematic boundaries, the engine naturally interpolates between continuous hydrodynamics in dense regions ($N \gtrsim 30$), stochastic kinetic relaxation in intermediate domains, and ballistic free-streaming in dilute regions ($N < 2$). This dynamical decoupling circumvents the need for an externally imposed Cooper-Frye particlization hypersurface. Instead, the kinetic decoupling scale, controlled locally by the oversampling parameter $\alpha$, allows the medium to transition seamlessly into a collisionless corona wherever the local expansion rate overwhelms the scattering rate.

With this conformal transport architecture validated, ShARK provides a foundation for modeling the pre-equilibrium and freeze-out dynamics of heavy-ion collisions. Planned extensions include implementing a non-conformal lattice QCD equation of state, enabling simulations with event-by-event $3+1$D fluctuating initial conditions. A continuous description of the fireball evolution from the initial collision to free-streaming, spanning the core-corona transition, would help clarify the microscopic origins of fluid behavior in the quark-gluon plasma.

The broader theoretical implications of this methodology, specifically, the ability to solve the exact relativistic Boltzmann equation without macroscopic gradient truncations, alongside a focused analysis of the universal non-equilibrium attractors and extreme radial profiles, are highlighted in the companion letter~\cite{ShARK_Letter}. By providing a continuous description of momentum transport from highly anisotropic initial states to dilute free-streaming limits, the ShARK framework is uniquely positioned to model far-from-equilibrium phenomena across disparate physical scales. Beyond heavy-ion physics, future applications will range from resolving extreme astrophysical environments, such as neutron star merger ringdowns, to modeling transport dynamics in early universe cosmology.

\section*{Acknowledgments}

The authors thank the members of the ExTrEMe Collaboration and R. Krupczak for fruitful discussions. The authors acknowledge the use of the Gemini AI model in code development. The authors are supported by CNPq through the INCT-FNA grant 408419/2024-5. T.N.dS. was supported by the Universal Grant 409029/2021-1. J.L.B. acknowledges a PhD fellowship from Fundacao de Amparo a Pesquisa e Inovacao do Estado de Santa Catarina (FAPESC) under grant 1947/2025. G.T. thanks Bolsa de produtividade CNPQ 305731/2023-8 and FAPESP temático 2023/13749--1 for support.

\bibliography{biblio}

\end{document}